\documentclass[12pt]{article}

\usepackage{graphicx,fancyhdr,bm,color,float}
\usepackage[bf]{caption}
\usepackage{subfig}
\usepackage{amsmath,amssymb,amsthm}
\usepackage{amsfonts}
\usepackage[onehalfspacing]{setspace}
\usepackage[colorlinks,pagebackref=false,linktoc=all]{hyperref}
\usepackage{natbib}
\usepackage[pdftex,letterpaper,left=1in,top=1in,bottom=1in,right=1in]{geometry}
\usepackage{comment}
\usepackage{booktabs}
\usepackage{adjustbox}
\usepackage{threeparttable}
\usepackage{adjustbox}

\definecolor{dark-blue}{rgb}{0.2,0.2,0.8}
\hypersetup{
   citecolor=dark-blue,
   bookmarksnumbered=true,
   urlcolor=black,
   linkcolor=dark-blue,
}

\begin{document}

\title{\bigskip\bigskip\bigskip
Trade Tensions Meet the Gig Economy: \\Evidence from the 2025 Halving of \\Chinese Travel to Japan\footnote{We are grateful to Daisuke Adachi, Daiji Kawaguchi, and all participants of the 29th Labor Economics Conference for helpful comments. This work has been supported by JST ERATO JPMJER2301 and JSPS KAKENHI 25K16620, Japan. We thank Timee, Inc. for providing the data and Takahide Kimura for his technical and institutional knowledge of the platform.}}
\author{
  Suguru Otani\thanks{\href{mailto:suguru.otani@e.u-tokyo.ac.jp}{suguru.otani@e.u-tokyo.ac.jp}, Market Design Center, Department of Economics, The University of Tokyo} \and
  Chang Yang\thanks{\href{mailto:yotyo@e.u-tokyo.ac.jp}{yotyo@e.u-tokyo.ac.jp}, Graduate School of Economics, The University of Tokyo}
}
\maketitle
\thispagestyle{empty}

\begin{abstract}
\noindent
How do geopolitical shocks to services exports reach local labor markets? Using China's November 2025 Japan travel advisory and postings on Japan's largest \mbox{spot-work} platform, we provide the first such evidence for the gig economy. Posted hourly wages fall immediately and persistently in prefectures more dependent on Chinese visitors, relative to others, whereas postings, hours, and employment show no robust change. Wages fall more for early morning and daytime shifts, when hotels and shops hire, than for evening and night shifts. Where each shift is priced anew, a \mbox{services-trade} shock lowers the price of labor in the hours it hits.

\medskip
\noindent\textit{JEL classification:} F14, F16, J21 \hfill 

\end{abstract}

\newpage

\section{Introduction}\label{sec:intro}

Geopolitical tensions have become a source of trade disruption in their own right, and much of the evidence on how they reach the economy concerns trade in goods \citep{gopinath2025changing}. Services exports have received less attention, although declines in these exports can directly affect local labor markets. Tourism services are consumed at the destination, and unused capacity cannot be stored. When visitors from abroad stop buying hotel stays, meals, and retail and transport services, demand for the workers who provide them falls. In this paper, we provide the first evidence on the short-run effects of such a loss on the gig segment of local labor markets, where work is hired shift by shift, and on whether the loss is absorbed through the price of labor or through the quantity of work.

The question is open because the evidence on labor markets comes from payroll records of ongoing employment relationships. \citet{kurmann2026neighbors} follow the employment, hours, and hourly wages of the workers of small U.S. businesses through the 2025 decline in Canadian visits and find that exposed establishments shed workers with no significant change in hours per worker or hourly wages. Such records show realized employment and pay, but not the price at which new work is offered, because they do not record the wage posted for a vacancy. Whether a loss of visitors is absorbed through the price of labor where jobs are created anew is therefore unknown. We answer the question using the travel advisory that China issued against Japan in November 2025 and confidential records from \textit{Timee}\footnote{The data provided by Timee, Inc. for this study have been processed and statistically aggregated in a manner that prevents the identification of any specific individuals. All data were provided and utilized in compliance with relevant Japanese laws and regulations (including the Act on the Protection of Personal Information), as well as Timee's Terms of Service and Privacy Policy.}, Japan's largest spot-work platform, on which businesses hire workers one shift at a time, at a wage posted anew for each shift, a form of platform-mediated gig work that has expanded rapidly worldwide \citep{boeri2020solo,kanayama2026nonparametric}. The records cover every posting on the platform through March 2026, with its location, occupation, hourly wage, hours, and shift start time.

On November 14, 2025, China's Ministry of Foreign Affairs and its embassy and consulates in Japan advised Chinese citizens to avoid traveling to Japan in the near future.\footnote{Embassy of the People's Republic of China in Japan, ``Reminder for Chinese Citizens to Avoid Travel to Japan in the Near Term'' (in Chinese), November 14, 2025, \url{https://jp.china-embassy.gov.cn/lsfw_0/lstx_138409/202511/t20251114_11753712.htm}.} The next day, major Chinese airlines began to allow free refunds and changes of tickets to Japan, through the end of December 2025, so that travelers could cancel their trips at once. Some airlines, notably the state-owned carriers, also canceled flights between China and Japan. We therefore date the shock at the week of the advisory.

\begin{figure}[!ht]
  \begin{center}
  \includegraphics[width = 0.8\textwidth]{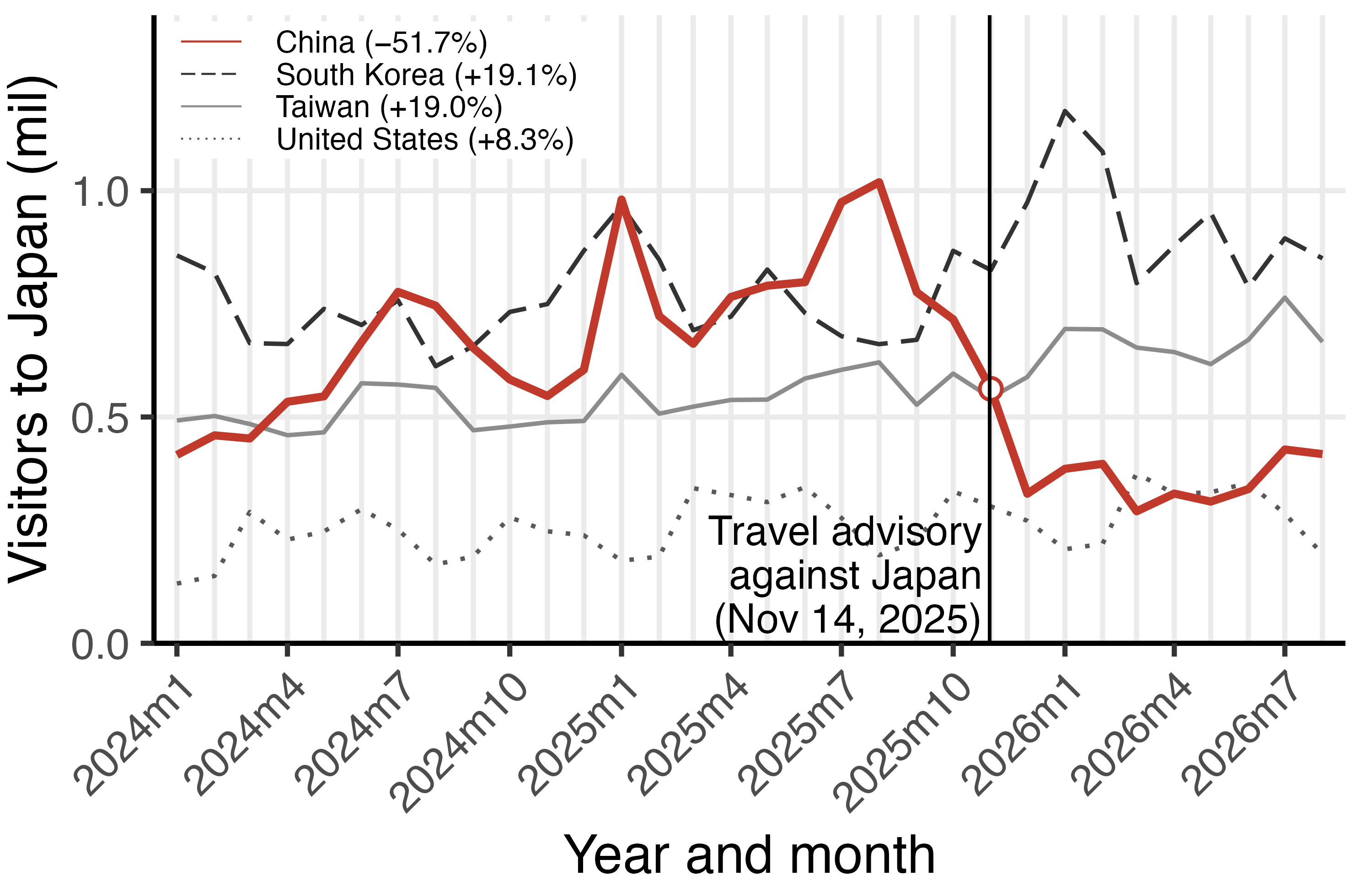}
  \caption{Inbound Visitors to Japan around the November 2025 Travel Advisory}
  \label{fg:china_monthly}
  \end{center}
  \footnotesize
  \textit{Notes}: Monthly visitors to Japan from China and from South Korea, Taiwan, and the United States, the three largest origins other than China. The hollow point is November 2025 for China, which includes the 13 days before the advisory. The percentage in the legend is the change in total visitors from November 2025 to August 2026 relative to the same months a year earlier.
\end{figure}

Figure~\ref{fg:china_monthly} plots monthly visitors to Japan from China and from South Korea, Taiwan, and the United States, the three largest other origins. Visits from China fell by 51.7 percent from November 2025 to August 2026 relative to the same months a year earlier, while visits from the other three origins rose by 8 to 19 percent. The shock is therefore specific to visitors from China and not part of a broader fall in inbound tourism demand. It is also very uneven across Japan: our Chinese tourism exposure measure was about 10 percent in Osaka and 8 percent in Kyoto in 2024, and under 1 percent in about half of the prefectures, and the prefectures with the higher shares lost the most Chinese stays after the advisory, as Section~\ref{sec:tourism} documents. This unevenness is the source of identifying variation in our design.

We focus on the spot labor market for two reasons. First, it is the margin on which the businesses that serve visitors adjust their staffing. Hotels, restaurants, and shops face demand that varies from day to day and from hour to hour and meet it with shifts that are posted, priced, and filled one at a time, alongside a regular staff whose employment is costly to adjust. A loss of visitors can therefore reduce demand for spot work quickly. The segment is also growing, and a shock can reach workers as fewer shifts or lower pay without necessarily producing a measured transition into unemployment. Second, each shift is priced anew, so the price of labor can adjust quickly, a margin that may be less flexible in ongoing employment relationships. The records also show whether each posting is filled, allowing us to distinguish the price offered, the quantity demanded, and realized employment by posting and shift start time.

Our design compares prefectures that differ in how dependent they were on Chinese visitors before the shock. The treatment intensity is the share of overnight stays by Chinese visitors in 2024 in each of the 47 prefectures. We aggregate the posting-level data to the prefecture-by-week level and estimate a continuous-dose difference-in-differences model with prefecture and week fixed effects and with controls for confounding factors that coincide with the shock. We then estimate an event study, which traces the dynamic effect after the advisory and checks whether trends were parallel before it.

We first estimate the effects on the posted wage, the number of postings, posted labor hours, and employment for all shifts. Posted wages fall in more exposed prefectures: a prefecture whose Chinese share is one percentage point higher posts wages about 0.27 percent lower after the shock, a gap that opens in the first four-week block and lasts to the end of the sample, and wages fall by about 1 percent in prefectures with a share of 2 percent or more, relative to those below 1 percent. The same specification at the establishment level gives a decline of 0.17 percent per percentage point with unit and week fixed effects, which absorb time-invariant differences across establishment--occupation units. Postings, posted labor hours, and employment fall in levels but show no significant change relative to the same week of the previous year. This wage response is consistent with the flexibility to set a new wage for each shift when expected revenue falls. In this market, a loss of visitors therefore reaches the price of labor at once. \citet{kurmann2026neighbors} find no significant change in the average hourly wage of employees, which they note could reflect changes in workforce composition or short-run wage rigidity; the posted wage of new work, which our data observe, is where such a shock can show up in the price of labor.

Decomposing the postings by the hour at which the shift starts provides evidence on the channels behind this decline. Because each shift is priced separately, wage declines can reflect both changes in labor supply across the prefecture and changes in labor demand at particular times of day. The pattern is consistent with the latter channel. For shifts that start in the early morning and the daytime, when most hotel and retail shifts begin, the posted wage falls by 0.27 and 0.31 percent per percentage point of the Chinese share, whereas the wages of evening and night shifts show no significant change, and among the early morning, daytime, and evening shifts, postings and posted labor hours fall most for early morning shifts, by 1.8 and 2.5 percent per percentage point. The larger wage declines in early morning and daytime are consistent with lower demand for labor in the hours in which hotels and shops hire. Together, the results indicate that a loss of visitors reaches spot workers through the price offered for the shifts that tourism demands, which the posting records reveal and records of standing employment do not.

\paragraph{Related literature.} Our paper relates to two strands of the literature. The first concerns the effects of trade and tourism shocks on local economies. Studies of trade shocks to local labor markets mostly concern imports of goods \citep{autor2013china}, and those of tourism mostly examine what visitors do to a place over years or decades \citep{faber2019tourism,hidalgo2024your,almagro2025location,matsuura2026impact}. By contrast, we study the short-run effects of an abrupt adverse tourism shock, whose timing was set by a decision of the Chinese government rather than by local conditions in Japan, which allows a cleaner identification of causal effects. \citet{gonzalez2025more} study short-run tourism shocks in Spain and find that employment gains in tourism and other services are offset by losses in other sectors. Political conflict has been shown to reduce trade in goods \citep{michaels2010freedom,heilmann2016political} and, through consumer boycotts, accommodation demand \citep{ahn2022political,greaney2025regional}. Shocks to Sino-Japanese relations in particular have been shown to lower the stock market value of Japanese firms exposed to China \citep{fisman2014nationalism}. We follow this type of shock, a fall in a trading partner's demand for the services that Japan exports, into the labor market, in the first months after an abrupt stop in visits from a single origin. The closest paper is \citet{kurmann2026neighbors}, who study the fall in Canadian visits to the United States using high-frequency employment, hours, and hourly wage records of small businesses and find employment losses in exposed markets, with no significant change in hours per worker or hourly wages. In a spot labor market, where the wage is set anew at every posting, we find instead that the posted wage falls at once, in the hours in which hotels and shops hire.

The second strand concerns spot labor markets, such as those intermediated by Uber, TaskRabbit, Instawork, and Timee, and the broader literature on alternative work arrangements \citep{katz2019rise,mas2020alternative,boeri2020solo,dube2020monopsony,angrist2021uber,cullen2021outsourcing,kanayama2026nonparametric}. Because both employers and workers in gig work can adjust instantly, posting and accepting shifts one at a time, these markets are well suited to studying short-run responses. Our wage results also relate to \citet{hazell2025downward}, who document downward rigidity in job-level wages posted for new hires. Our paper contributes to this literature by showing how a spot labor market adjusts when it is hit by a shock from outside, a fall in international tourism.

\section{Data}\label{sec:data}

\subsection{Tourism Data}\label{sec:tourism}

We measure exposure to Chinese tourism at the level of Japan's 47 prefectures. The visitor counts in Figure~\ref{fg:china_monthly} are monthly arrivals by origin from the Japan National Tourism Organization. The exposure measure comes from the Overnight Travel Statistics Survey of the Japan Tourism Agency, which reports overnight stays by the guest's nationality or region of origin. Following \citet{kurmann2026neighbors}, for each prefecture $p$, we define
\begin{equation}
  \text{Chinese share}_{p} = \frac{\text{Overnight stays by Chinese visitors}_{p,2024}}{\text{Total overnight stays}_{p,2024}},
\end{equation}
computed from totals over 2024, the last full year before the shock, and we define the shares of Korean, Taiwanese, and American visitors, the three largest origins other than China by number of arrivals, in the same way. The nationality-specific numerator covers facilities with at least 10 employees; the denominator includes facilities of all sizes. Because the measure is fixed before the shock, it proxies for each prefecture's structural dependence on Chinese tourism, not a post-shock outcome.\footnote{The prefecture is the finest level at which the overnight stays by visitor origin are published. Under some assumptions, the Chinese share can also be computed for 130 wide-area municipal regions, each a group of neighboring municipalities, which divide the 47 prefectures into about three regions each on average. Appendix~\ref{app:region} reports estimates at this level.}

\begin{figure}[!htbp]
  \begin{center}
  \subfloat[Chinese share of overnight stays]{\includegraphics[width = 0.49\textwidth]{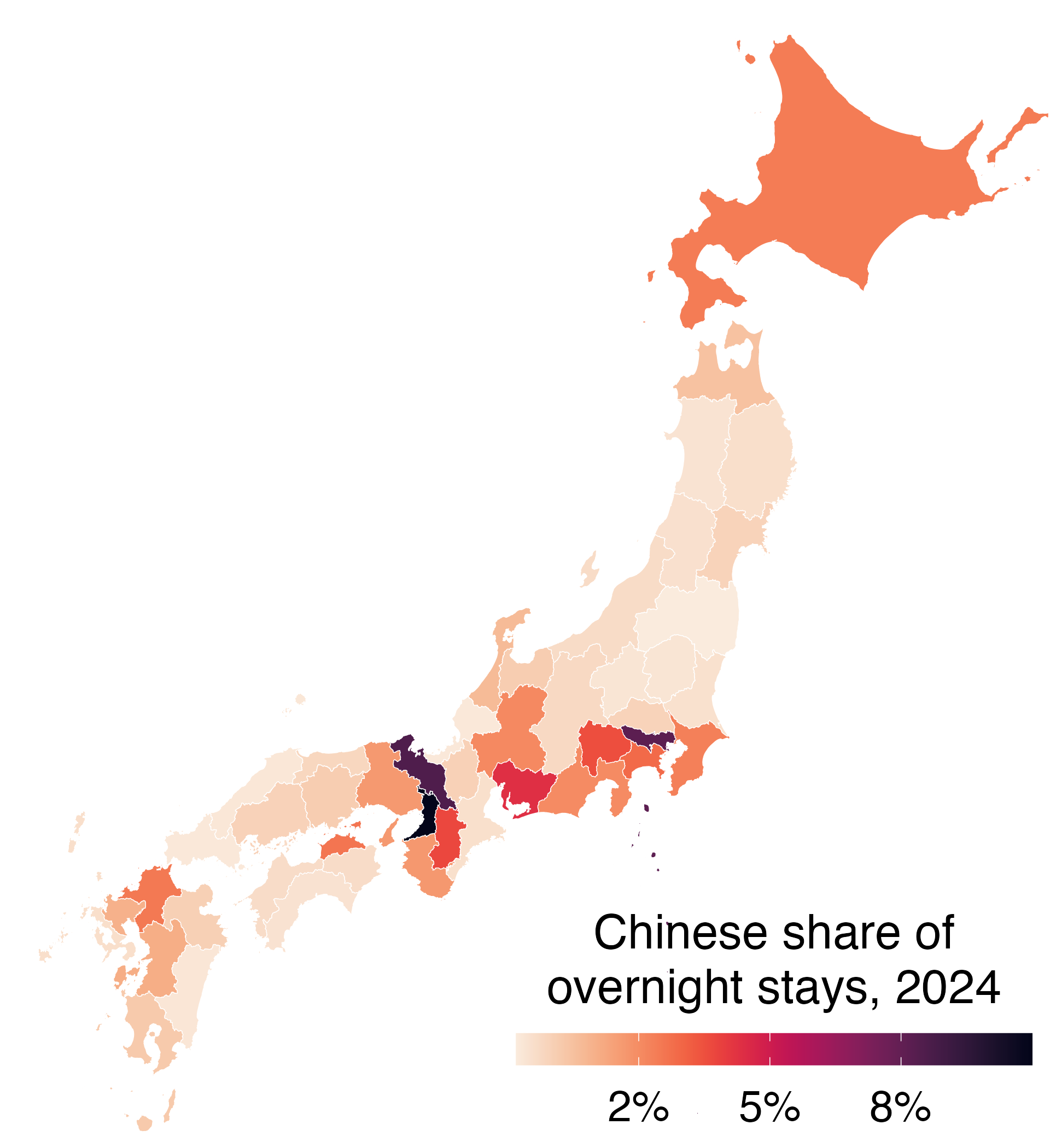}}\hfill
  \subfloat[Spot-work postings in hotels, restaurants, and retail]{\includegraphics[width = 0.49\textwidth]{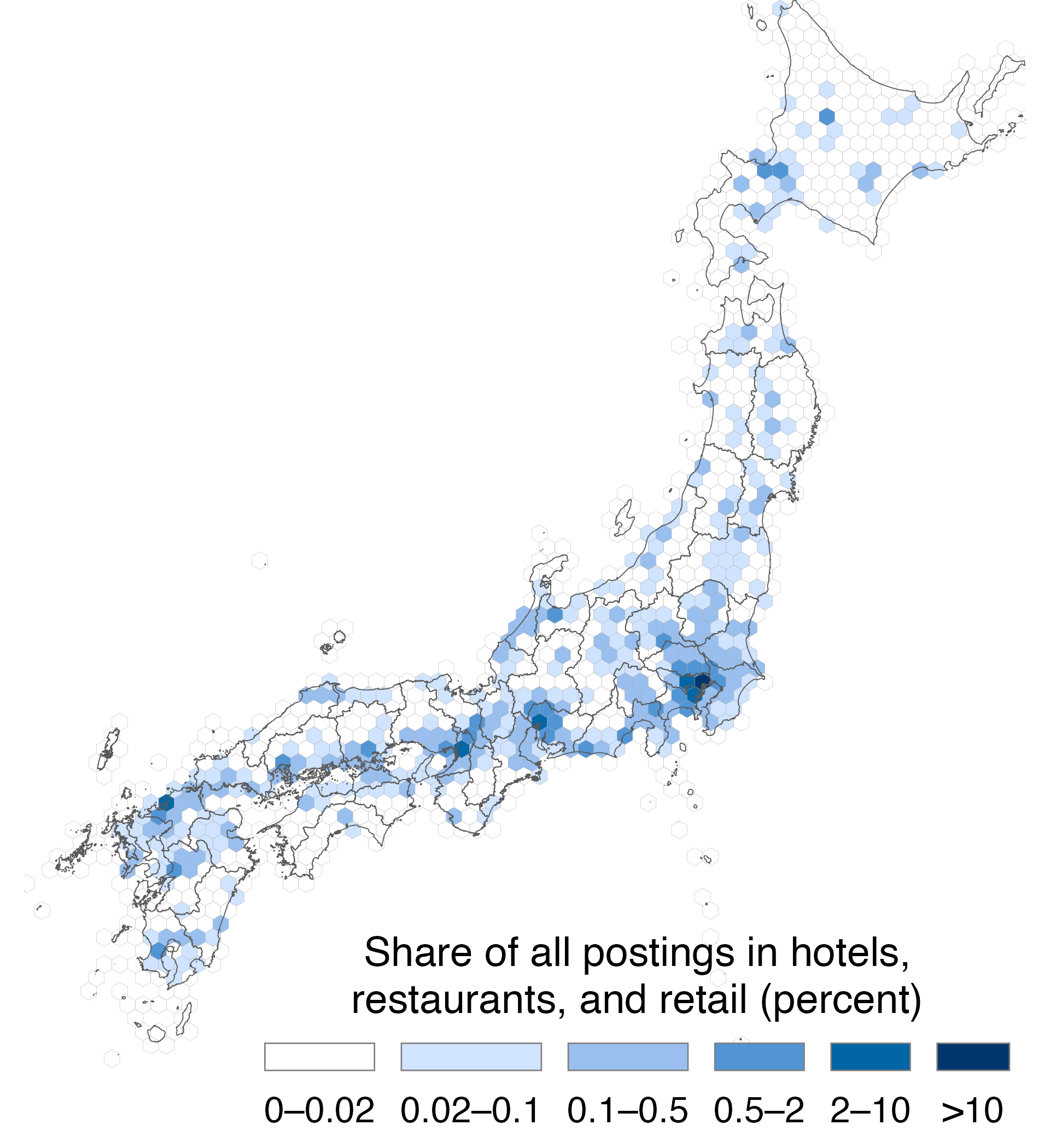}}\\
  \subfloat[Share and the change in Chinese overnight stays]{\includegraphics[width = 0.49\textwidth]{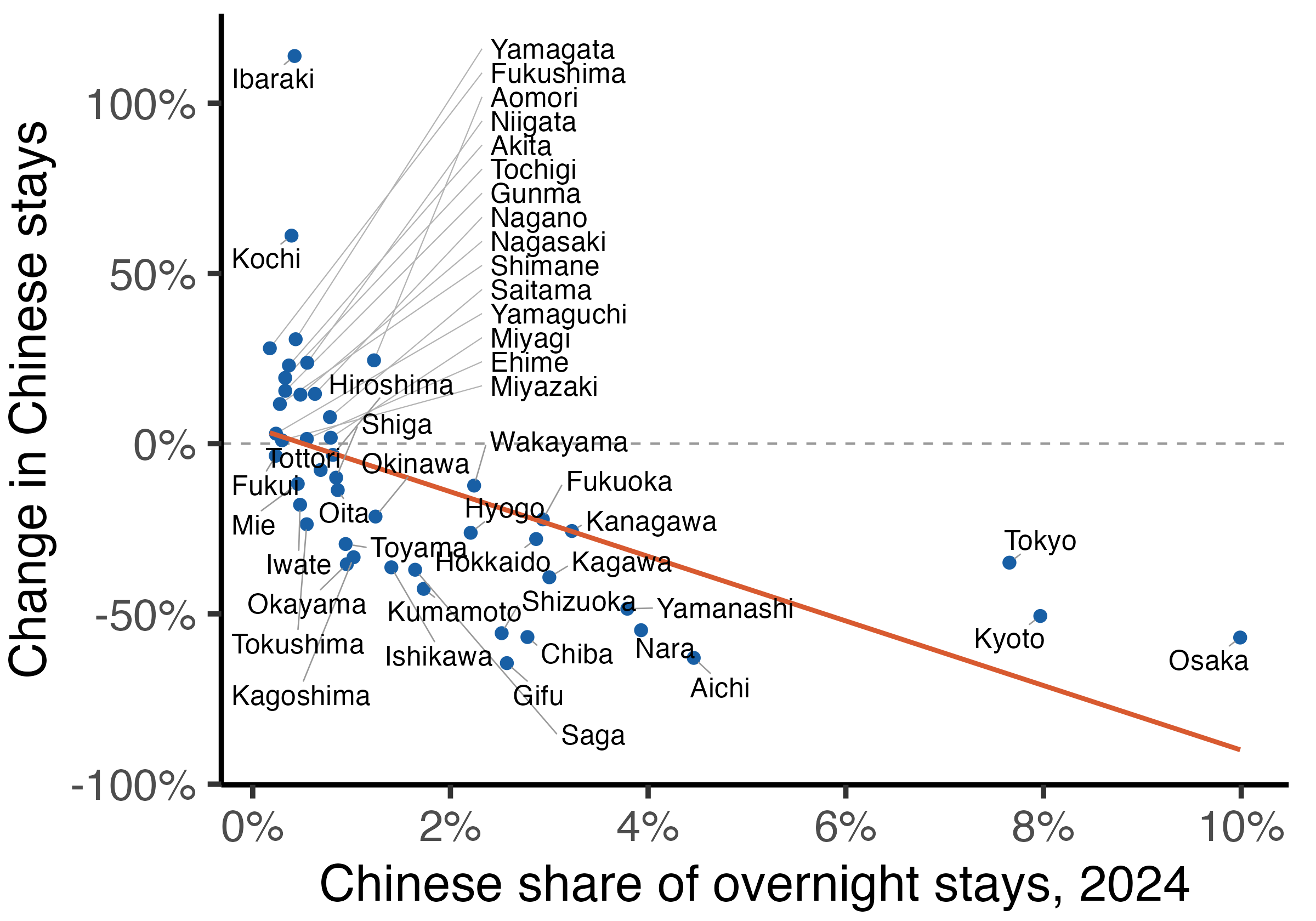}}\hfill
  \subfloat[Shift start time by occupation group]{\includegraphics[width = 0.49\textwidth]{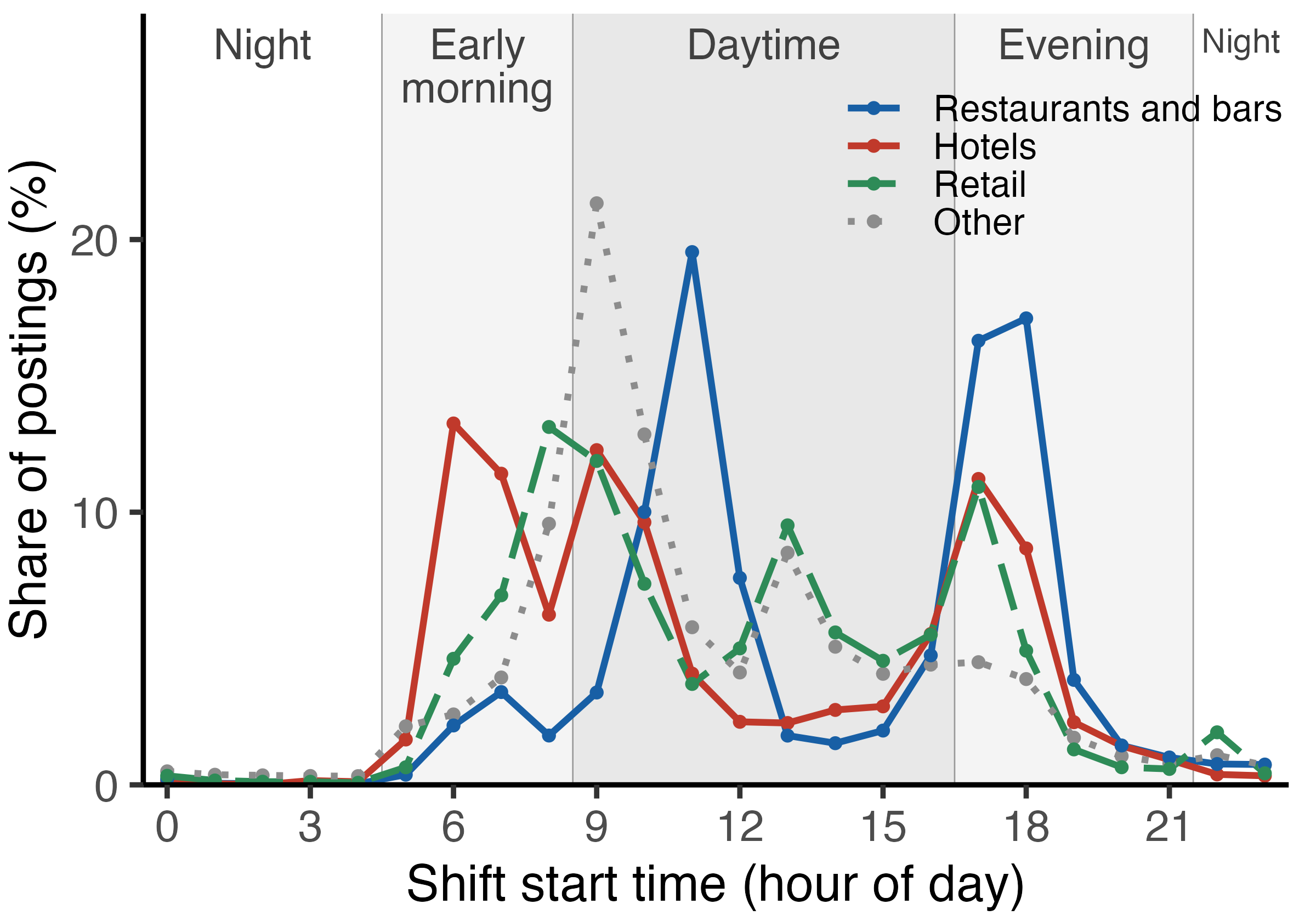}}
  \caption{Exposure to Chinese Tourism and the Granularity of the Spot-Work Data}
  \label{fg:China_visitor_dependency_map}
  \end{center}
  \footnotesize
  \textit{Notes}: Panel (a): Chinese share of overnight stays by prefecture, calculated from 2024 Chinese stays at facilities with at least 10 employees divided by total stays at facilities of all sizes. Panel (b): share of all spot-work postings in hotels, restaurants and bars, and retail from November 2024 to the week of November 3, 2025, on hexagons 25 kilometers across. Panel (c): Chinese share against the change in Chinese overnight stays from November 2025 to July 2026 relative to the same months a year earlier, one point per prefecture. Panel (d): share of each occupation group's postings by shift start hour; the shaded bands are the shift start-time bands of Section~\ref{sec:shift_band}. Appendix~ \ref{app:spot} gives the definitions.
\end{figure}

Figure~\ref{fg:China_visitor_dependency_map} shows the measure and checks that it captures exposure to the shock. Panel (a) maps the Chinese share of overnight stays across prefectures.\footnote{The map in panel (a) of Figure~\ref{fg:China_visitor_dependency_map}, like the map in panel (b) described in Section~\ref{sec:spotwork}, omits Okinawa and the remote islands, which lie far outside the frame of the main islands. Okinawa's Chinese share is 1.24 percent, the 19th highest, as Table~\ref{tab:share_china_pref} reports, and all estimates use the 47 prefectures.} Panel (c) plots, for each prefecture, the change in Chinese overnight stays from November 2025 to July 2026 relative to the same months a year earlier against this share. Chinese stays fell most in the prefectures with the highest shares, by about 57 percent in Osaka, 51 percent in Kyoto, and 35 percent in Tokyo, while many prefectures with shares near zero saw no decline, and the fitted line implies a fall of about 10 percentage points for each percentage point of additional share. The measure therefore separates prefectures that lost many Chinese visitors from those that lost few.\footnote{Table~\ref{tab:stays_did} in Appendix~\ref{app:tourism} estimates this relation within prefecture-by-origin-by-month cells. In a prefecture with a one percentage point higher Chinese share, the growth of Chinese overnight stays after the advisory is 9.9 percent lower relative to other origins, whereas that of guests from Taiwan, Korea, and the United States, relative to the remaining origins, is 2.9 percent higher.} Exposure is concentrated: the median prefecture has a Chinese share of 0.9 percent and the 90th percentile 3.8 percent, while Osaka (10.0 percent), Kyoto (8.0 percent), and Tokyo (7.7 percent) are far above the rest.

\subsection{Spot-Work Data}\label{sec:spotwork}

We use the term ``spot-work platform'' for \textit{Timee}, the leading online platform in Japan's spot-work market \citep{kanayama2026nonparametric,kanayama2026justminimumwagehikes,sekiya2026designing}. Spot work means employment contracts, subject to Japanese labor law including the minimum wage, for short, one-off engagements that are arranged through a digital platform, often for the same day or the near term. Timee connects businesses with workers through a mobile app. Workers can fill shifts in food service, retail, and logistics, among other industries.

On the labor-demand side, businesses must first pass a screening and verification process before they are allowed to post. Once registered, they create a posting by specifying the tasks, the wage, and the shift times, and the platform makes the posting immediately visible to workers and matches them with the business, which employs them directly. On the labor-supply side, workers register for free and apply to available shifts without a traditional hiring process such as interviews or long-term contracts. Some jobs require qualifications, such as nursing or childcare, but many postings require no formal qualifications or prior experience. The platform earns a commission on each completed match rather than charging for postings or subscriptions.

Our labor-market data are confidential posting- and contract-level records from Timee, covering job postings (vacancies) and realized matches for jobs located across Japan, in all occupations. For each posting we observe the establishment's location and occupation, the posted hourly wage, the posted shift length and number of workers, and precise timestamps for when the posting appears and, if filled, when it is matched. We aggregate the postings to the prefecture-by-week level and construct four outcome variables: the mean posted hourly wage, weighted by posted labor hours (the posted headcount multiplied by the shift length) so that it reflects the wage of the typical hour offered, the number of postings, total posted labor hours, and employment, the number of completed matches of the postings made in the week.\footnote{We drop the 0.1 percent of postings with more than 25 workers, the 99.9th percentile of the posted headcount. Appendix~\ref{app:spot} gives further detail on the data.} For the robustness checks, we also aggregate the postings to 130 wide-area municipal regions by week and to establishment-by-occupation units by week.

Panels (b) and (d) of Figure~\ref{fg:China_visitor_dependency_map} show the granularity of the posting records in space and in time of day. Panel (b) places the postings in hotels, restaurants and bars, and retail, the industries that serve visitors, on hexagons 25 kilometers across by the location of the establishment. The postings are concentrated in the large cities: the hexagon centered on Tokyo Station is the only one with more than 10 percent of these postings, and the hexagons in the Tokyo, Osaka, Nagoya, and Fukuoka areas hold 2 to 10 percent each, while postings also appear in every region. Panel (d) plots the distribution of shift start times by occupation group. Hotel shifts concentrate in the early morning, whereas restaurant shifts concentrate at 11:00 and in the evening.

Our baseline sample runs from the week of November 4, 2024, to the week of March 23, 2026, in the 47 prefectures.\footnote{The year-over-year analysis in Table~\ref{tab:main_did} uses data back to November 2023 as the base year.} The first fully treated week is the week of November 17, 2025 (relative week 0). The week of November 10, which contains the advisory date (Friday, November 14), is partially treated and is dropped. The pre-shock period therefore covers 53 weeks, about one year, and the post-shock period 19 weeks, so that aggregating the individual postings to the prefecture-by-week level gives 3,384 prefecture-weeks.

\section{Empirical Strategy}\label{sec:strategy}

We estimate a continuous-dose difference-in-differences specification at the prefecture-week level as the baseline:
\begin{equation}
  \begin{split}
  \log(y_{p,t}) ={}& \beta \left(\text{Post}_t \times \text{Chinese share}_{p}\right)
  + X_{p,t}'\gamma + \delta_p + \lambda_t + \varepsilon_{p,t},
  \end{split}
  \label{eq:main}
\end{equation}
where $y_{p,t}$ is an outcome (posted hourly wage, number of postings, total posted labor hours, or employment), constructed from the postings of all occupations, in prefecture $p$ in week $t$; $\text{Post}_t$ equals one from the week of November 17, 2025; and $\delta_p$ and $\lambda_t$ are prefecture and week fixed effects. All outcomes are in logs and estimated by OLS. Standard errors are clustered by prefecture.

The vector of controls $X_{p,t}$ contains three terms for changes that coincide with the advisory. The first is the 2024 shares of overnight stays by Korean, Taiwanese, and American visitors, $\text{Share}_{c,p}$ for $c \in \{\text{Korea, Taiwan, USA}\}$, each interacted with $\text{Post}_t$, which allow post-advisory changes to vary with pre-shock exposure to these origins.\footnote{These controls do not necessarily remove substitution toward visitors from other origins.} The second is the log of the prefectural minimum wage. Prefectural minimum wages are revised every autumn, and the 2025 revisions took effect in the same period as the advisory. The third is the interaction $\text{Kinki}_p \times \text{Expo}_t$, where $\text{Kinki}_p$ marks the seven prefectures of the Kinki region\footnote{Mie, Shiga, Kyoto, Osaka, Hyogo, Nara, and Wakayama.} and $\text{Expo}_t$ equals one from the week of April 14 to the week of October 6, 2025, the weeks of the World Expo in Osaka (April 13 to October 13, 2025). The Expo was a one-time event that created labor demand in the Kinki region for the months before the advisory and ended shortly before it. Because Osaka and Kyoto, two of the prefectures with the highest Chinese shares, are in the Kinki region, the end of the Expo could otherwise be mistaken for an effect of the advisory, so we control for it separately.

To trace the dynamics and test for pre-trends, we estimate the event-study specification
\begin{equation}
  \log(y_{p,t}) = \text{Chinese share}_{p} \times \sum_{\tau \neq -1} \beta_\tau \mathbf{1}[\tau(t) = \tau]
  + X_{p,t}'\gamma + \delta_p + \lambda_t + \varepsilon_{p,t},
  \label{eq:event}
\end{equation}
where $\beta_\tau$ is the average effect in the $\tau$-th four-week block after the advisory. The block of week $t$ is $\tau(t) = \lfloor w_t / 4 \rfloor$, where $w_t$ is the number of weeks since the week of November 17, 2025, so that $\tau = 0$ covers the first four weeks of the advisory and $\tau = -1$, the omitted reference, the weeks just before it.\footnote{The equation is estimated with weekly data, so the fixed effects $\lambda_t$ vary by week as in equation~\eqref{eq:main}; the indicator $\mathbf{1}[\tau(t) = \tau]$ equals one for every week in block $\tau$. We pool the coefficients into four-week blocks to reduce noise. The reference block consists of three weeks, October 20 to November 9, 2025, because the week of November 10, which contains the advisory, is dropped. In $X_{p,t}$, the other-origin shares are interacted with the block indicators instead of $\text{Post}_t$.} The coefficients $\beta_\tau$ for $\tau \geq 0$ trace how the effect evolves after the advisory, and we report them with 90 percent confidence intervals.

To see at which times of the day the wage falls, we estimate equations~\eqref{eq:main} and~\eqref{eq:event} separately by the start time of the shift. We split the postings into four bands by the start time: early morning (5:00 to 8:59), daytime (9:00 to 16:59), evening (17:00 to 21:59), and night (22:00 to 4:59). We aggregate the postings of each band to the prefecture-week, with the outcomes defined as in the main panel, and estimate the specification with the same controls and fixed effects on each band panel.

The identifying assumption is that, conditional on the fixed effects and controls, prefectures with different exposure would have evolved in parallel in the absence of the advisory, and that no other change correlated with exposure coincided with it. Under this assumption the coefficients $\beta_\tau$ for $\tau < -1$ are zero, which the event study examines, and column (2) of Table~\ref{tab:main_did} compares each outcome with the same week of the previous year.

\section{Results}\label{sec:results}

\subsection{Difference-in-Differences Estimates}\label{sec:static}

Table~\ref{tab:main_did} reports the main result in column (1), which uses the postings of all occupations aggregated to the prefecture level.\footnote{Appendix~\ref{app:region} reports the analysis at the level of the 130 wide-area municipal regions and of establishments. Appendix~\ref{app:by_spec} shows that the estimate is stable when the minimum wage and the Expo indicator are added to the controls for other origins. When the specification also controls for the timing of the 2025 minimum-wage revisions, with dummies for the weeks since each prefecture's revision, the wage estimate falls to $-0.15$ but remains significant, as column (5) of Table~\ref{tab:main_did_by_spec} shows. We keep the specification without these dummies as the baseline because many revisions took effect in the weeks around the advisory, so the dummies also absorb part of the change after the advisory.} A prefecture whose Chinese share is one percentage point higher posts wages 0.27 percent lower after the advisory. The difference is sizable: Table~\ref{tab:bins_did} in Appendix~\ref{app:functional} shows that wages fall by about 1 percent in prefectures with a share of 2 percent or more, relative to those below 1 percent.

The wage response is consistent with flexibility in newly posted wages. Terms are set for each posting, subject to the minimum wage, so an establishment whose shifts bring in less revenue can offer less at its next posting; the gap opens in the first four-week block after the advisory, as Section~\ref{sec:es} shows.

\begin{table}[!htbp]
  \begin{center}
      \caption{Difference-in-Differences Estimates: All Shifts and by Shift Start Time}
      \label{tab:main_did}
      \resizebox{\textwidth}{!}{\begin{tabular}[t]{lcccccc}
\toprule
\multicolumn{1}{c}{ } & \multicolumn{2}{c}{All shifts} & \multicolumn{4}{c}{By shift start time} \\
\cmidrule(l{3pt}r{3pt}){2-3} \cmidrule(l{3pt}r{3pt}){4-7}
\multicolumn{1}{c}{ } & \multicolumn{1}{c}{Levels} & \multicolumn{1}{c}{Year over year} & \multicolumn{1}{c}{Early morning} & \multicolumn{1}{c}{Daytime} & \multicolumn{1}{c}{Evening} & \multicolumn{1}{c}{Night} \\
\cmidrule(l{3pt}r{3pt}){2-2} \cmidrule(l{3pt}r{3pt}){3-3} \cmidrule(l{3pt}r{3pt}){4-4} \cmidrule(l{3pt}r{3pt}){5-5} \cmidrule(l{3pt}r{3pt}){6-6} \cmidrule(l{3pt}r{3pt}){7-7}
 & (1) & (2) & (3) & (4) & (5) & (6)\\
\midrule
\emph{Posted hourly wage (log)} &  &  &  &  &  & \\
Share $\times$ Post & $-0.272^{***}$ & $-0.350^{***}$ & $-0.269^{***}$ & $-0.310^{***}$ & $-0.167^{}$ & $-0.180^{}$\\
 & (0.076) & (0.107) & (0.094) & (0.083) & (0.102) & (0.146)\\
\emph{Number of postings (log)} &  &  &  &  &  & \\
Share $\times$ Post & $-1.180^{*}$ & $-1.086^{}$ & $-1.818^{**}$ & $-1.003^{}$ & $-0.635^{}$ & $-4.652^{***}$\\
 & (0.593) & (0.891) & (0.866) & (0.635) & (0.785) & (1.523)\\
\emph{Posted labor hours (log)} &  &  &  &  &  & \\
Share $\times$ Post & $-1.763^{***}$ & $-1.031^{}$ & $-2.507^{**}$ & $-1.432^{**}$ & $-1.429^{**}$ & $-5.900^{***}$\\
 & (0.633) & (0.943) & (1.230) & (0.686) & (0.693) & (2.040)\\
\emph{Employment (log)} &  &  &  &  &  & \\
Share $\times$ Post & $-1.331^{**}$ & $0.041^{}$ & $-1.802^{}$ & $-1.068^{*}$ & $-0.999^{}$ & $-5.852^{**}$\\
 & (0.563) & (0.666) & (1.172) & (0.565) & (0.880) & (2.422)\\
\midrule
Prefecture FE & Yes & -- & Yes & Yes & Yes & Yes\\
Week FE & Yes & Yes & Yes & Yes & Yes & Yes\\
Controls & Yes & Yes & Yes & Yes & Yes & Yes\\
Pre-shock share of postings (\%) & 100.0 & 100.0 & 17.3 & 58.7 & 21.0 & 3.0\\
Observations & 3,384 & 3,384 & 3,384 & 3,384 & 3,384 & 3,364\\
Observations, employment & 3,102 & 1,692 & 3,102 & 3,102 & 3,102 & 3,076\\
\bottomrule
\end{tabular}
}
  \end{center}\footnotesize
  \textit{Notes}: Each coefficient is from a separate regression of the log of the outcome on Share $\times$ Post, where Share is the prefecture's 2024 Chinese share of overnight stays and Post equals one from the week of November 17, 2025. The unit is the prefecture-week (47 prefectures). In column (2), the outcome is the log difference from the same week of the previous year and the controls enter as changes from the previous year. Columns (3) to (6) use the postings whose shift starts in the band of the column, defined in Section~\ref{sec:strategy}. Employment is the number of completed matches of the postings made in the week and uses the postings made up to the week of February 9, 2026; in column (2), it uses the weeks from June 2, 2025, the first for which the same week of the previous year is observed. The wage is the labor-hours-weighted mean posted hourly wage. All columns include the controls of Section~\ref{sec:strategy}. Pre-shock share is the band's share of postings from November 2024 to the week of November 3, 2025. Standard errors clustered by prefecture in parentheses. *, **, *** denote significance at 10, 5, 1 percent.
\end{table}

The numbers of postings and posted labor hours point in the same direction. In column (1), postings and posted labor hours are 1.2 and 1.8 percent lower per percentage point of the Chinese share, as a fall in the demand for visitor-facing work would imply. Column (2) compares each week with the same week of the previous year, which removes seasonality and a constant trend; Figure~\ref{fg:event_study} shows that postings and posted labor hours were already falling in more exposed prefectures before the advisory. The year-over-year estimates in column (2) show no significant change. The fourth block reports employment, the number of completed matches of the postings made in a week, for the postings made up to the week of February 9, 2026, the last week for which completed matches are observed. Employment is 1.3 percent lower per percentage point in column (1) and shows no significant change in column (2). Employment thus moves with postings. The wage estimate of $-0.35$ in column (2) remains significant and close to that of column (1). The wage falls in both columns, whereas the decline in postings, posted labor hours, and employment depends on whether seasonality is removed.

\subsection{Event Study}\label{sec:es}

\begin{figure}[!ht]
  \begin{center}
  \subfloat[Posted hourly wage (log)]{\includegraphics[width = 0.49\textwidth]{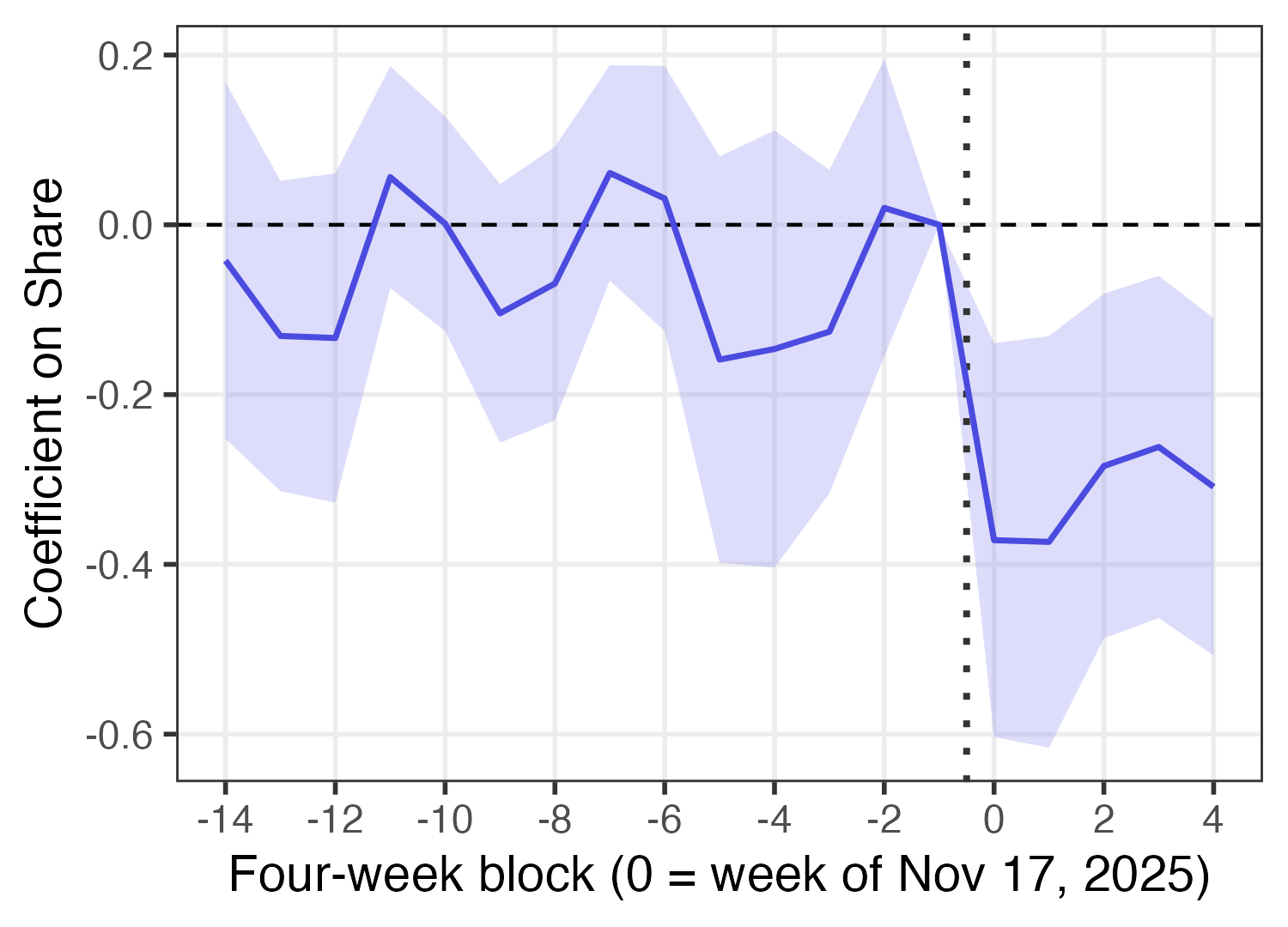}}\hfill
  \subfloat[Number of postings (log)]{\includegraphics[width = 0.49\textwidth]{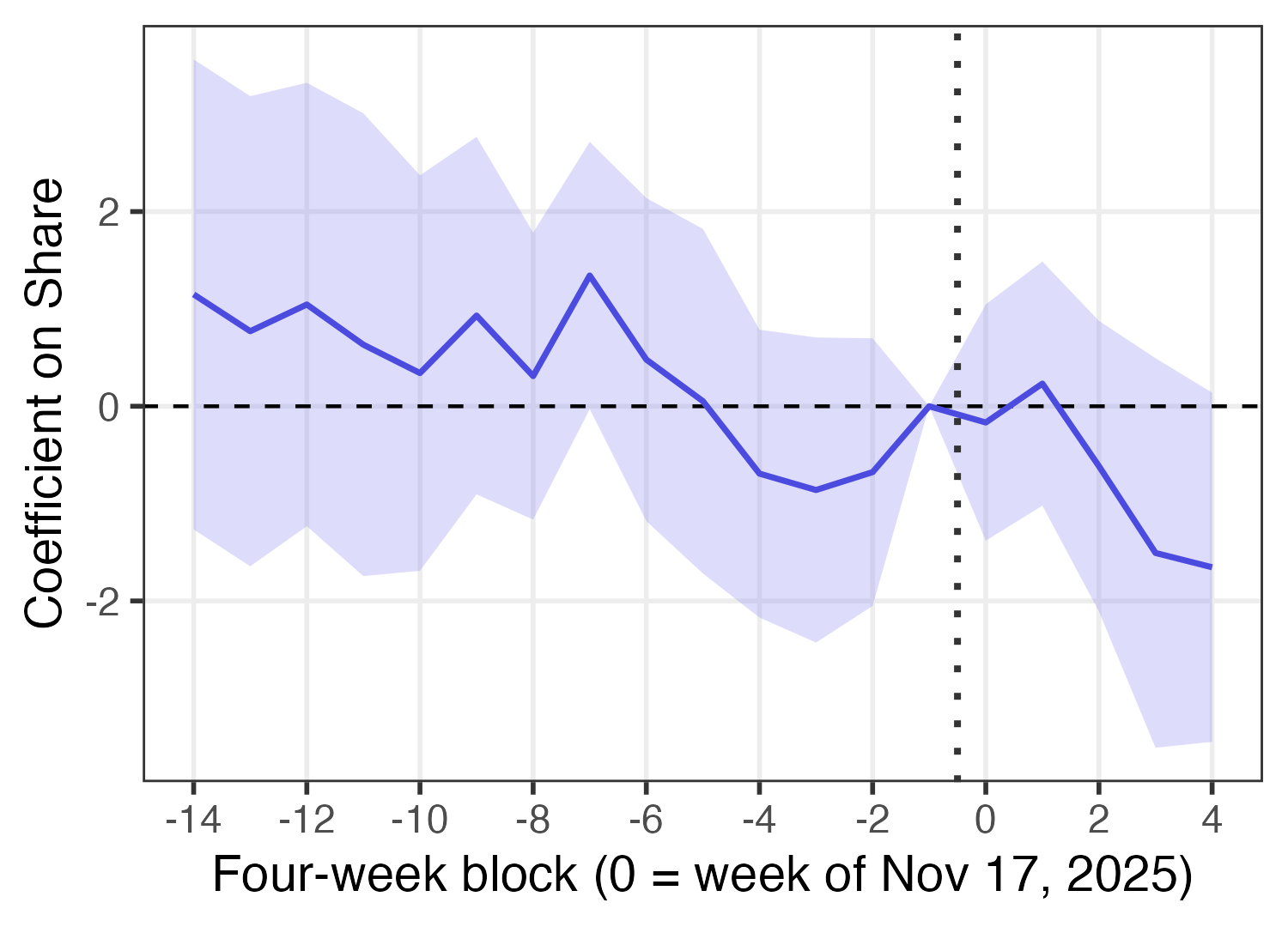}}\\
  \subfloat[Posted labor hours (log)]{\includegraphics[width = 0.49\textwidth]{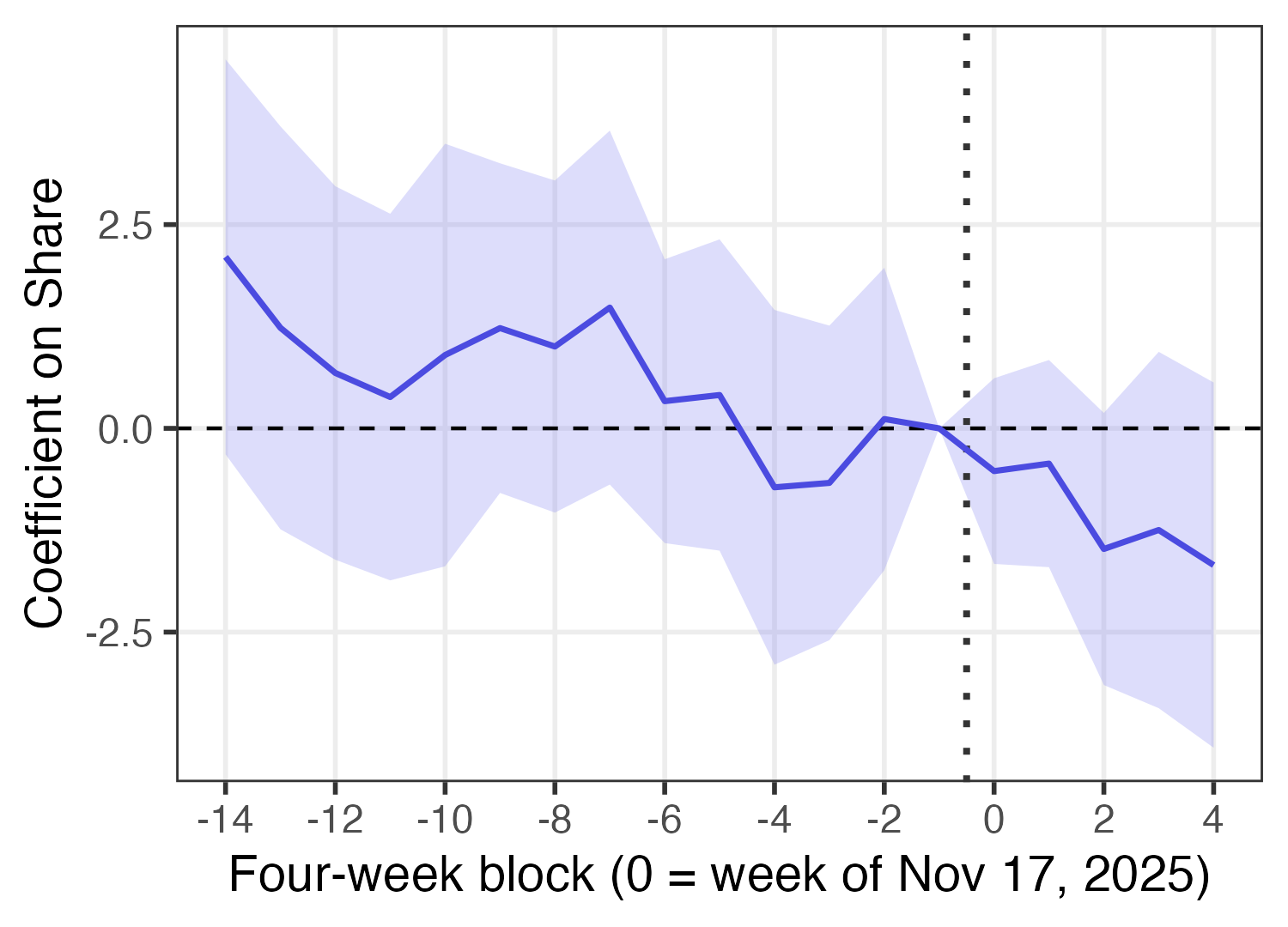}}\hfill
  \subfloat[Employment (log)]{\includegraphics[width = 0.49\textwidth]{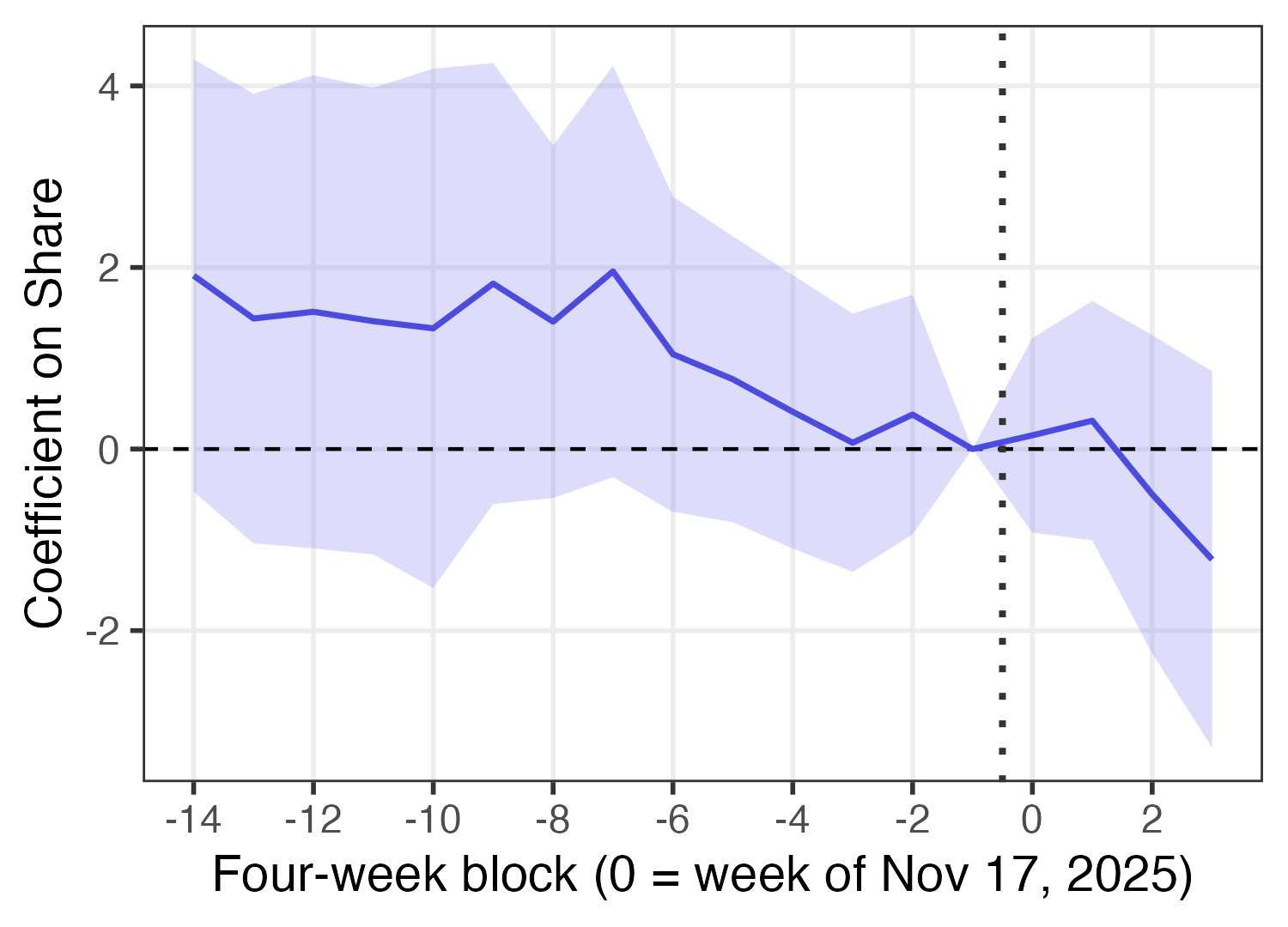}}
  \caption{Event Study: Posted Wage, Postings, Posted Labor Hours, and Employment}
  \label{fg:event_study}
  \end{center}
  \footnotesize
  \textit{Notes}: Coefficients on Share (the prefecture's 2024 Chinese share of overnight stays) interacted with four-week blocks relative to the week of November 17, 2025, from column (1) of Table~\ref{tab:main_did}. The outcomes are the logs of the posted hourly wage (Panel (a)), postings (Panel (b)), posted labor hours (Panel (c)), and employment, the number of completed matches of the postings made in the week (Panel (d)); Panel (d) uses the postings made up to the week of February 9, 2026. The shaded area is the 90 percent confidence interval, and the block immediately before the shock is the omitted category. Standard errors are clustered by prefecture.
\end{figure}

Figure~\ref{fg:event_study} estimates the effect separately for each four-week block, which shows whether the gap between more and less exposed prefectures opened at the advisory or was already there. Panel (a) shows the posted wage. The pre-advisory coefficients are small and show no trend. In the first four weeks after the advisory the coefficient falls to $-0.37$, and it remains between $-0.26$ and $-0.37$ through the end of the sample.
For postings, posted labor hours, and employment in Panels (b) to (d), the coefficients after the advisory show no significant change. Figure~\ref{fg:event_study_yoy} in Appendix~\ref{app:by_spec} reports the year-over-year event study.

\subsection{Additional Analysis}\label{sec:shift_band}

\paragraph{Heterogeneity by Shift Start Time.} Tourism creates demand for labor at particular times of the day. Hotels post shifts for housekeeping, breakfast service, and check-out that start in the early morning and the daytime, and shops, attractions, and transport serve visitors during the day, whereas restaurants and bars also serve residents in the evening. Panel (d) of Figure~\ref{fg:China_visitor_dependency_map} shows this pattern in the postings: hotel shifts concentrate in the early morning, and restaurant shifts in the evening. Differences in wage responses across these hours therefore indicate whether the decline reflects tourism-related demand.

Before the shock, daytime shifts account for 59 percent of postings, early morning shifts for 17 percent, evening shifts for 21 percent, and night shifts for 3 percent, as Table~\ref{tab:main_did} reports. The four band panels add up to the panel of column (1).

Columns (3) to (6) of Table~\ref{tab:main_did} report the estimates by band next to the estimates for all shifts in column (1). The posted wage falls by 0.27 percent per percentage point of the Chinese share for early morning shifts and by 0.31 percent for daytime shifts, in line with the decline of 0.27 percent for all shifts. For evening and night shifts there is no significant change. The wage decline is therefore larger for the shifts that start in the early morning and the daytime, the hours in which hotels and shops post most of their shifts, than for the evening and night shifts. The table also reports postings and posted labor hours by band. Among the early morning, daytime, and evening shifts, both fall most for early morning shifts, by 1.8 and 2.5 percent per percentage point; for daytime and evening shifts, posted labor hours fall by 1.4 percent and postings show no significant change. Employment falls significantly for night shifts and, at the 10 percent level, for daytime shifts, and shows no significant change for early morning and evening shifts. The night coefficients are larger, but night shifts are 3 percent of postings. Appendix~\ref{app:band_diff} compares the wages of early morning and daytime shifts with those of evening and night shifts within the same prefecture and week, which absorbs shocks common to the whole prefecture, such as the minimum-wage revisions; when the bands are weighted by their posted labor hours, the wage of early morning and daytime shifts falls significantly more than that of evening and night shifts.

\begin{figure}[!ht]
  \begin{center}
  \includegraphics[width = 0.95\textwidth]{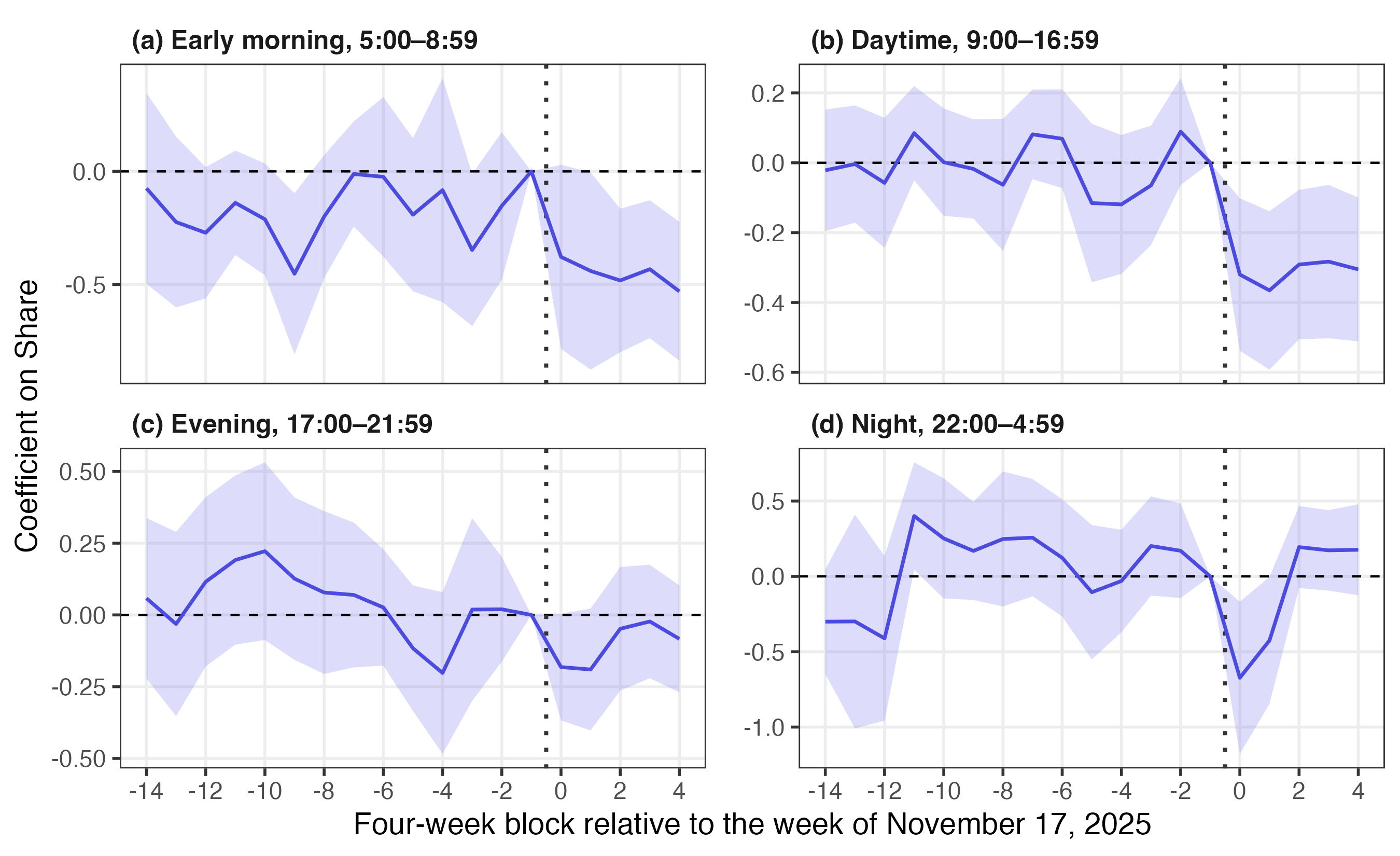}
  \caption{Event Study by Shift Start Time: Posted Wage}
  \label{fg:event_study_shift_band}
  \end{center}
  \footnotesize
  \textit{Notes}: Coefficients on Share interacted with four-week blocks relative to the week of November 17, 2025, from column (1) of Table~\ref{tab:main_did}, estimated separately on the postings whose shift starts in each band. The outcome is the log of the band's labor-hours-weighted mean posted hourly wage. The shaded area is the 90 percent confidence interval, and the block immediately before the shock is the omitted category. Standard errors are clustered by prefecture.
\end{figure}

Figure~\ref{fg:event_study_shift_band} shows the event study for the posted wage in each band. In Panels (a) and (b), the wage of early morning and daytime shifts steps down in the first four-week block after the shock and stays down through the end of the sample, the profile of the aggregate wage in Figure~\ref{fg:event_study}, and the declines in these two bands are large. In Panel (c), the evening coefficients fall in the first two blocks after the shock and return toward zero thereafter, and in Panel (d), the night coefficients are volatile both before and after the shock. The persistent wage decline is more pronounced for early morning and daytime shifts.

\paragraph{Heterogeneity by Occupation.} The platform records the occupation of each posting, and Figure~\ref{fg:industry} in Appendix~\ref{app:industry} estimates the main specification separately for the 15 occupations with the most postings. The posted wage falls significantly for restaurant staff and for convenience stores, supermarkets, and drugstores, two of the occupations that serve visitors most directly, whereas the decline for hotels is not significant. At the 5 percent level, postings and posted labor hours fall significantly for restaurant staff and show no significant change for hotels or for convenience stores, supermarkets, and drugstores. The wage decline is thus present in the occupations that serve visitors, not only in the aggregate.

\paragraph{Worker Mobility.} Appendix~\ref{app:worker} examines whether incumbent workers in exposed prefectures work fewer hours or work farther from home after the advisory. Incumbent workers are those who completed at least three shifts between November 2024 and April 2025, in any occupation. The hours per incumbent and the share of incumbents who worked show no significant change, whereas the median commuting distance and the share of hours worked outside the prefecture of residence rise when the specification also controls for the timing of the minimum-wage revisions.

\subsection{Robustness}\label{sec:robust}

\paragraph{Establishment-level estimates.} Because the wage falls in the first four-week block, it could reflect a change in which jobs are posted rather than lower wages for the same jobs. Appendix~\ref{app:estab} therefore compares each establishment with itself before and after the advisory, and the posted wage still falls, by 0.17 percent per percentage point of the Chinese share.

\paragraph{Dose-response and placebo test.} Appendix~\ref{app:functional} shows that the wage estimate does not depend on the prefectures with the highest shares. Dropping Osaka, Kyoto, and Tokyo, one at a time or together, leaves the coefficient between $-0.26$ and $-0.34$, and bins of the share show a decline in the prefectures with shares of 2 percent or more, but no larger one above 3 percent. Because there are only 47 clusters, Appendix~\ref{app:placebo} also assesses significance by placebo. We re-estimate the model 500 times on the pre-shock weeks, each time with a randomly chosen fake shock date and with the Chinese shares randomly reassigned across prefectures. None of the 500 placebo wage coefficients is as large in absolute value as the actual estimate.

\paragraph{Wage response to the realized loss of Chinese stays.} Appendix~\ref{app:iv} uses the share as an instrument for the realized loss of Chinese stays. A prefecture with a one percentage point higher share loses 0.77 percentage points more of its total stays, a strong first stage, as Table~\ref{tab:iv_realized} shows. A loss of Chinese stays equal to one percent of a prefecture's total stays lowers the posted wage by 0.35 percent.

\section{Conclusion}\label{sec:concl}

We provide the first evidence on how a loss of services exports reaches local spot labor markets, using China's Japan travel advisory in November 2025 and the posting records of Japan's largest spot-work platform. Posted wages fall, relative to other prefectures, in the prefectures that depended more on Chinese visitors, immediately after the advisory and persistently, whereas postings, posted labor hours, and employment show no robust change. Decomposing the postings by the hour at which the shift starts, we find larger wage declines in the hours in which hotels and shops hire, consistent with a reduction in tourism-related labor demand. This contrasts with \citet{kurmann2026neighbors}, who find that the fall in Canadian visits to the United States reduces employment in exposed markets with no significant change in hourly wages. Where each shift is priced anew, a services-trade shock reaches the posted wage at once, in the hours in which the affected businesses hire. More generally, where the price of new work adjusts, geopolitical shocks to trade can leave employment unchanged and reach workers through pay, a margin that assessments tracking only employment miss.





\newpage
\bibliographystyle{ecca}
\bibliography{utmd_timee}

\clearpage
\appendix

\begin{center}
{\Large\bfseries Supplementary Appendix}\\[1ex]
{\large Trade Tensions Meet the Gig Economy: Evidence from the 2025 Halving of Chinese Travel to Japan}
\end{center}

\bigskip

\renewcommand{\thetable}{A\arabic{table}}
\renewcommand{\thefigure}{A\arabic{figure}}
\setcounter{table}{0}
\setcounter{figure}{0}

\section{Tourism Data}\label{app:tourism}

\paragraph{Construction of the exposure measure.} The Overnight Travel Statistics Survey of the Japan Tourism Agency reports overnight stays at accommodation facilities by prefecture, month, and the origin of the guest. For each prefecture, the Chinese share is Chinese overnight stays at facilities with at least 10 employees in 2024 divided by total overnight stays at facilities of all sizes in that year, as in the equation of Section~\ref{sec:tourism}. We fix the measure at 2024, the last full year before the advisory. The survey reports stays by origin only for facilities with at least 10 employees, so the numerator excludes Chinese guests at smaller facilities while the denominator includes all guests. The measure therefore understates the Chinese share of stays.

\paragraph{The measure by prefecture.} Table~\ref{tab:share_china_pref} lists the Chinese share for all 47 prefectures from the highest to the lowest. Osaka (10.0 percent), Kyoto (8.0 percent), and Tokyo (7.7 percent) are far above the rest, the median prefecture has a share of 0.9 percent, and the 90th percentile is 3.8 percent.

\begin{table}[!htbp]
  \begin{center}
      \caption{Chinese Share of Overnight Stays by Prefecture, 2024}
      \label{tab:share_china_pref}
      \begin{tabular}{rlrrlr}
\toprule
Rank & Prefecture & Share (\%) & Rank & Prefecture & Share (\%) \\
\midrule
1 & Osaka & 9.99 & 25 & Shiga & 0.84 \\
2 & Kyoto & 7.97 & 26 & Hiroshima & 0.81 \\
3 & Tokyo & 7.65 & 27 & Miyagi & 0.79 \\
4 & Aichi & 4.46 & 28 & Saitama & 0.78 \\
5 & Nara & 3.93 & 29 & Tottori & 0.69 \\
6 & Yamanashi & 3.79 & 30 & Nagano & 0.63 \\
7 & Kanagawa & 3.23 & 31 & Niigata & 0.55 \\
8 & Kagawa & 3.00 & 32 & Ehime & 0.55 \\
9 & Fukuoka & 2.94 & 33 & Tokushima & 0.55 \\
10 & Hokkaido & 2.87 & 34 & Nagasaki & 0.48 \\
11 & Chiba & 2.78 & 35 & Iwate & 0.48 \\
12 & Gifu & 2.57 & 36 & Mie & 0.46 \\
13 & Shizuoka & 2.52 & 37 & Yamagata & 0.43 \\
14 & Wakayama & 2.24 & 38 & Ibaraki & 0.42 \\
15 & Hyogo & 2.20 & 39 & Kochi & 0.39 \\
16 & Kumamoto & 1.73 & 40 & Akita & 0.37 \\
17 & Saga & 1.64 & 41 & Gunma & 0.33 \\
18 & Ishikawa & 1.40 & 42 & Tochigi & 0.33 \\
19 & Okinawa & 1.24 & 43 & Miyazaki & 0.30 \\
20 & Aomori & 1.23 & 44 & Shimane & 0.27 \\
21 & Kagoshima & 1.02 & 45 & Yamaguchi & 0.24 \\
22 & Okayama & 0.95 & 46 & Fukui & 0.23 \\
23 & Toyama & 0.94 & 47 & Fukushima & 0.17 \\
24 & Oita & 0.86 & &  &  \\
\bottomrule
\end{tabular}

  \end{center}\footnotesize
  \textit{Notes}: The Chinese share is the ratio of overnight stays by guests classified as Chinese in the survey at facilities with at least 10 employees to total overnight stays at facilities of all sizes, with both counts summed over 2024. Prefectures are ordered from the highest to the lowest share.
\end{table}

\begin{figure}[!ht]
  \begin{center}
  \includegraphics[width = 0.8\textwidth]{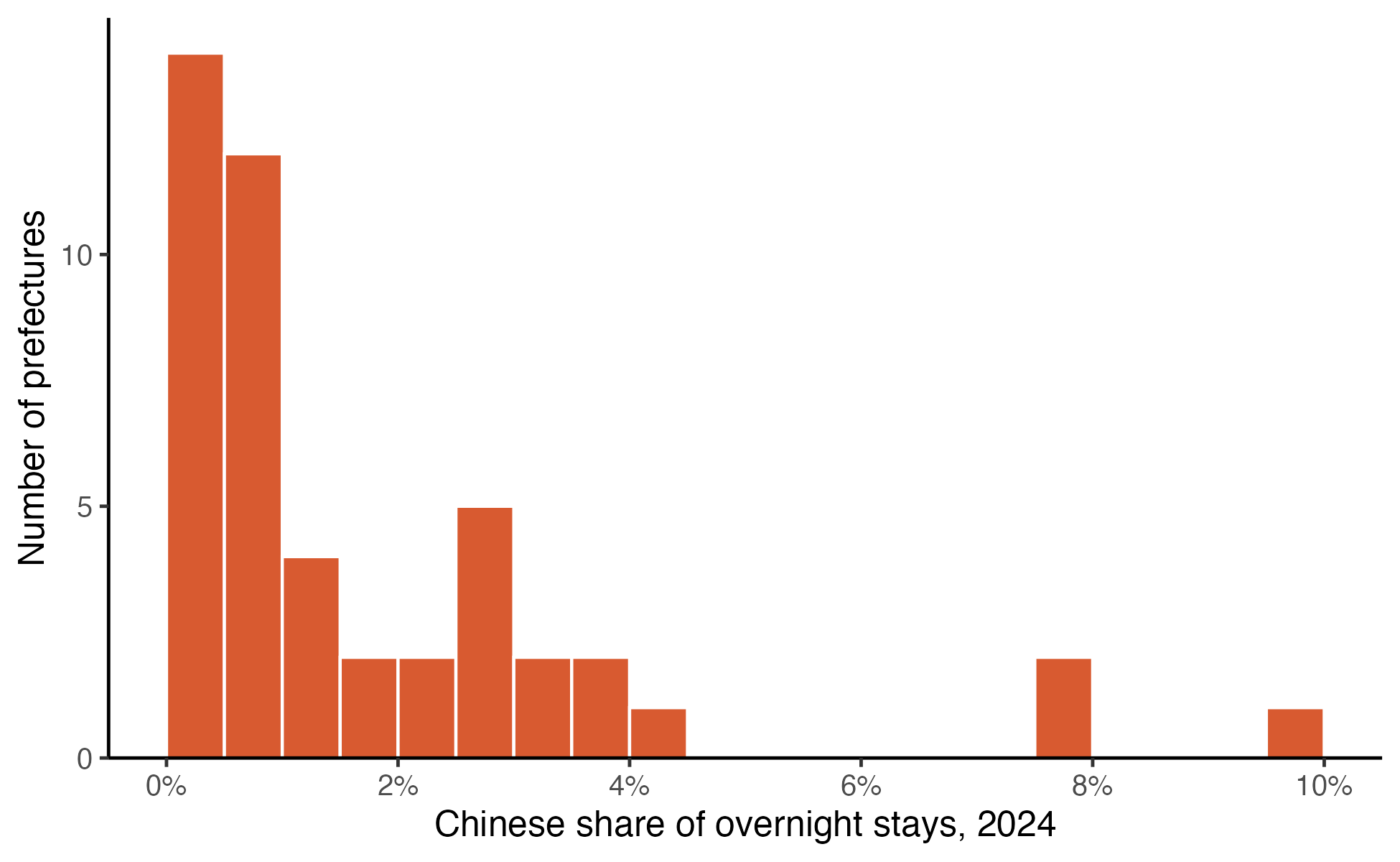}
  \caption{Distribution of the Chinese Share of Overnight Stays across Prefectures, 2024}
  \label{fg:share_hist}
  \end{center}
  \footnotesize
  \textit{Notes}: Histogram of the Chinese share of overnight stays in 2024 for the 47 prefectures, in bins of 0.5 percentage points.
\end{figure}

\paragraph{Distribution.} Figure~\ref{fg:share_hist} shows the distribution of the share across prefectures. It is skewed to the right: about half of the prefectures have a share below 1 percent, and three prefectures have a share above 7 percent, with the remaining prefectures spread in between.

\paragraph{Overnight stays by origin after the advisory.} In the months after the advisory, overnight stays of Chinese guests fell relative to the same months a year earlier, while those of guests from Taiwan, South Korea, and the United States rose. These comparisons, here and in Panel (c) of Figure~\ref{fg:China_visitor_dependency_map}, use nationality-specific counts that cover facilities with at least 10 employees before 2026. The 2026 figures by origin multiply stays at facilities with at least 20 rooms by the prefecture-month ratio of foreign stays at all facilities to foreign stays at those facilities. This assumes the same origin composition across facility sizes and gives the 2026 series broader facility coverage than the earlier counts. The survey also changed its stratification criterion from employee counts to room counts in January 2026, so year-over-year comparisons may partly reflect these changes. If prefectures that depend on Chinese guests also attract more guests from these origins, the decline in the labor market could be mitigated by them, and the main specification controls for the shares of Taiwan, Korea, and the United States interacted with Post. Table~\ref{tab:stays_did} checks this with a triple-difference design in the spirit of \citet{ahn2022political}, which compares origins within a prefecture and month. The outcome is the growth of overnight stays from the same month of the previous year in prefecture-by-origin-by-month cells, with fixed effects for prefecture by origin, prefecture by month, and origin by month, and the Kinki-by-Expo control interacted with the origin group. Column (1) shows that Chinese stays fall more than those of other origins in prefectures with a higher Chinese share: a prefecture with a share one percentage point higher has a 9.9 percent larger decline in the growth of Chinese stays. Column (2) compares guests from Taiwan, Korea, and the United States together with the other origins except China: the growth of their stays is 2.9 percent higher in a prefecture with a share one percentage point higher, so that these guests fill part of the places left by Chinese guests in the prefectures that depended on them, which supports controlling for the shares of these origins.

\begin{table}[!htbp]
  \begin{center}
      \caption{Overnight Stays by Origin and the Chinese Share}
      \label{tab:stays_did}
      \begin{tabular}[t]{lcc}
\toprule
\multicolumn{1}{c}{ } & \multicolumn{2}{c}{Outcome: log growth of overnight stays, year over year} \\
\cmidrule(l{3pt}r{3pt}){2-3}
 & (1) & (2)\\
\midrule
Origin $j$ & China & Taiwan, Korea, and USA\\
\midrule
Share $\times$ 1[origin $= j$] $\times$ Post & $-9.917^{***}$ & $2.874^{***}$\\
 & (2.945) & (1.030)\\
\midrule
Comparison origins & All other origins & Other non-Chinese origins\\
Control & Yes & Yes\\
Prefecture $\times$ origin FE & Yes & Yes\\
Prefecture $\times$ month FE & Yes & Yes\\
Origin $\times$ month FE & Yes & Yes\\
Observations & 19,381 & 18,394\\
\bottomrule
\end{tabular}

  \end{center}\footnotesize
  \textit{Notes}: The unit is a prefecture-by-origin-by-month cell and the outcome is the log growth of overnight stays from the same month of the previous year. Each column reports the coefficient on Share $\times$ 1[origin $= j$] $\times$ Post for the origin $j$ in the heading, where Share is the prefecture's 2024 Chinese share of overnight stays and Post equals one from November 2025. Column (1) compares China with all other origins; column (2) compares Taiwan, Korea, and the United States with the remaining origins other than China. The control is the Kinki-by-Expo indicator interacted with origin $j$, and the fixed effects are prefecture by origin, prefecture by month, and origin by month. Standard errors clustered by prefecture in parentheses. *, **, *** denote significance at 10, 5, 1 percent. The 2026 values by origin are approximations based on facilities with 20 or more rooms. The scaling assumption and survey redesign are described in Appendix~\ref{app:tourism}.
\end{table}

\section{Spot-Work Data}\label{app:spot}

\paragraph{Platform and postings.} \textit{Timee} is a spot-work platform on which establishments post short shifts that workers fill without a standing employment relationship. Establishments pass a screening process before they can post; a posting specifies the task, the hourly wage, and the shift time, and is visible to workers immediately. Workers register for free and apply to shifts directly. The platform charges a commission on completed matches, not for postings.

\paragraph{Posting-level records.} For each posting we observe the location and occupation of the establishment, the posted hourly wage, the posted number of workers and shift length, and the timestamps at which the posting appears and, if filled, is matched. The records cover postings in all occupations across Japan. We drop the 0.1 percent of postings with more than 25 workers, the 99.9th percentile of the posted headcount; the longest posted shift is 16 hours, so a cap on shift length of 24 hours drops nothing.

\paragraph{Panels (b) and (d) of Figure~\ref{fg:China_visitor_dependency_map}.} The occupation groups are hotels; restaurants and bars (restaurant staff, izakaya, fast food, cafes, and bars); retail (convenience stores, supermarkets, and drugstores, apparel and general goods, in-store light work, mobile phone sales, and in-store food sampling); and all other occupations. The three groups that serve visitors account for half of the postings. Panel (b) uses the postings from November 2024 to the week of November 3, 2025, located by the address of the establishment; 92 percent of the postings have a geocoded address. The hexagons are 25 kilometers across on an equal-area projection centered on Japan, and the grid is centered on Tokyo Station, so that the hexagon containing Tokyo Station does not depend on the plotted area; Okinawa and the remote islands are outside the plotted area, as in panel (a), and each hexagon's share is relative to the postings placed on the map. Panel (d) uses the same postings and the hour at which the shift starts, and its bands are those of Section~\ref{sec:shift_band}.

\paragraph{Prefecture-by-week panel.} We assign each posting to the prefecture of its establishment and to the week, starting on Monday, in which it is posted, and aggregate the postings to the prefecture-by-week level. The number of postings is a count, and total posted labor hours is the sum of the posted headcount multiplied by the shift length. The posted wage is the mean hourly wage weighted by posted labor hours, so that it reflects the wage of the typical hour offered. The panel has 47 prefectures and 72 weeks, 3,384 prefecture-weeks, and every prefecture-week has at least one posting, so the logs of all three outcomes are defined.

\paragraph{Sample period.} The baseline sample runs from the week of November 4, 2024, to the week of March 23, 2026. The first fully treated week is the week of November 17, 2025; the week of November 10, which contains the advisory date, is partially treated and dropped, which leaves 72 weeks. The year-over-year estimates compare each week with the same week 364 days earlier, and use data back to the week of November 6, 2023 as the base year.

\paragraph{Establishment-by-occupation panel.} For the establishment-level estimates we build a panel of establishment-by-occupation units by week. We keep units with at least four posting weeks before the shock and follow them to the end of the sample, coding weeks without postings as zero for the counts; about 63 percent of the unit-weeks have no posting. The wage is observed only in weeks with at least one posting. For the year-over-year estimates, we keep units with at least four posting weeks in each of the two years before the shock, so that the sample is again defined by pre-shock information.

\paragraph{Summary statistics.} Table~\ref{tab:summary_stats} reports summary statistics of the posted hourly wage, the shift length, and the number of positions at the level of the posting (Panel (a)), and of the weighted wage, the number of postings, and posted labor hours at the level of the prefecture-by-week panel (Panel (b)). The 22.3 million postings have a mean posted wage of 1,150 yen, a mean shift length of 4.5 hours, and 1.6 positions on average (median 1). The prefecture-by-week cells are large but uneven: the mean cell has 6,601 postings and 51,217 posted labor hours, the median 3,083 postings, and the smallest cell has 415 postings, so no cell is thin.

\begin{table}[!htbp]
  \begin{center}
      \caption{Summary Statistics}
      \label{tab:summary_stats}
      \resizebox{\textwidth}{!}{\begin{tabular}[t]{lrrrrrr}
\toprule
 & N & Mean & Median & SD & Min & Max\\
\midrule
\multicolumn{7}{l}{\textit{(a) Posting level}}\\
\hspace{1em}Hourly wage (yen) & 22,337,340 & 1,149.52 & 1,141.00 & 128.89 & 951.00 & 2,000.00\\
\hspace{1em}Working hours & 22,337,340 & 4.53 & 4.00 & 1.93 & 1.00 & 16.00\\
\hspace{1em}Number of positions & 22,337,340 & 1.60 & 1.00 & 1.75 & 1.00 & 25.00\\
\multicolumn{7}{l}{\textit{(b) Prefecture-week level}}\\
\hspace{1em}Hourly wage, labor-hours weighted (yen) & 3,384 & 1,084.73 & 1,076.29 & 59.19 & 979.69 & 1,298.25\\
\hspace{1em}Number of postings & 3,384 & 6,600.87 & 3,082.50 & 9,404.45 & 415.00 & 85,993.00\\
\hspace{1em}Posted labor hours & 3,384 & 51,216.52 & 22,433.75 & 68,900.81 & 2,532.50 & 514,856.53\\
\bottomrule
\end{tabular}
}
  \end{center}\footnotesize
  \textit{Notes}: Number of observations, mean, median, standard deviation, minimum, and maximum of each variable from November 2024 to the week of March 23, 2026. Panel (a) is at the posting level, and the number of positions is the posted headcount. Panel (b) is at the prefecture-by-week level (47 prefectures); the wage is the mean posted hourly wage weighted by posted labor hours, and posted labor hours is the posted headcount times the shift length. Postings with more than 25 positions or shifts longer than 24 hours are excluded, and the week of November 10, 2025 is dropped.
\end{table}

\section{Additional Analysis}\label{app:robust}

\subsection{Estimates for Regions and Establishments}\label{app:region}\label{app:estab}

We repeat the analysis at two levels that do not rely on the prefecture aggregate. Table~\ref{tab:region_estab_did} reports the estimates; the Level columns use the main specification and the Year over year columns take account of seasonality as in column (2) of Table~\ref{tab:main_did}.

\begin{figure}[!ht]
  \begin{center}
  \includegraphics[width = 0.8\textwidth]{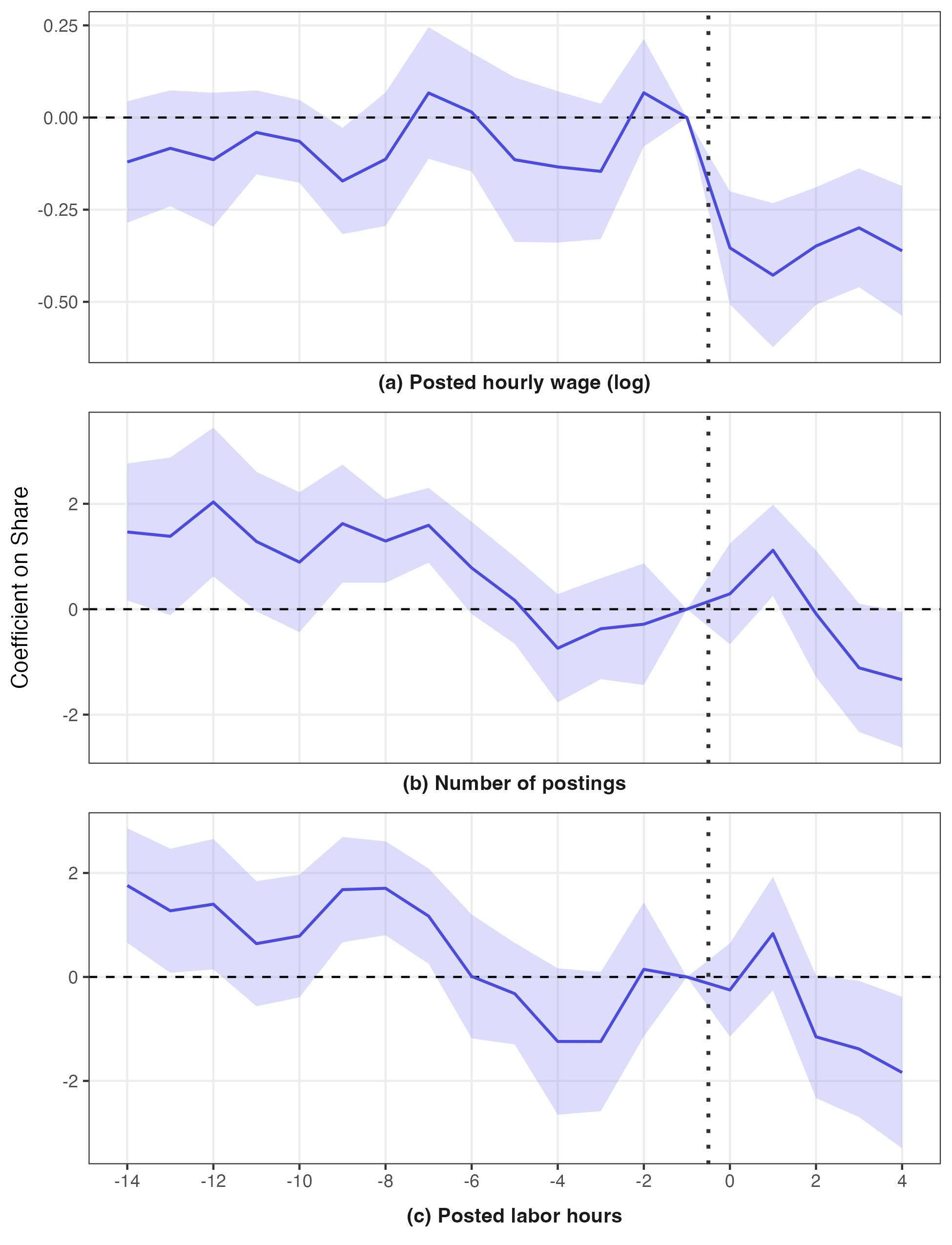}
  \caption{Event Study: Region Level}
  \label{fg:event_study_region130}
  \end{center}
  \footnotesize
  \textit{Notes}: Panels show the posted hourly wage, postings, and posted labor hours for column (1) of Table~\ref{tab:region_estab_did}. The shaded area is the 90 percent confidence interval. Standard errors are clustered by region.
\end{figure}

The first level is the 130 wide-area municipal regions, a finer geography than the prefecture, with the region's 2024 Chinese share of overnight stays as the exposure measure. The Overnight Travel Statistics Survey reports total and foreign overnight stays for each region but not their breakdown by origin, which it publishes only by prefecture. We therefore assume that, within a prefecture, the share of each origin in foreign overnight stays is the same in all of its regions and equal to the prefecture's value. For each month of 2024, the Chinese share of region $r$ in prefecture $p$ is then
\begin{equation}
  \text{Chinese share}_{r} = \frac{\text{Foreign overnight stays}_{r}}{\text{Total overnight stays}_{r}} \times \frac{\text{Overnight stays by Chinese visitors}_{p}}{\text{Foreign overnight stays}_{p}},
\end{equation}
and we average it over the twelve months. The shares of Korea, Taiwan, and the United States used as controls are built in the same way. The wage result holds at the finer geography: with the controls of the main specification, the coefficient is $-0.29$ in column (1), close to the prefecture-level estimate of $-0.27$. The event study in Figure~\ref{fg:event_study_region130} has the same profile as the prefecture-level one: the wage coefficients are small before the advisory, between about $-0.17$ and $+0.07$, and fall to about $-0.35$ to $-0.43$ in the first two four-week blocks and stay near $-0.3$ to $-0.4$ afterwards. For postings and posted labor hours, estimated by Poisson pseudo-maximum likelihood, the coefficients in column (1) are negative and significant at $-1.0$ and $-1.3$. The event study shows, however, that these two outcomes decline steadily relative to exposure from the start of the sample, so, as at the prefecture level, the quantity estimates cannot be separated from a trend that began before the advisory. The Year over year column (2) uses only the region-weeks with postings in both years. The wage coefficient is $-0.33$, close to column (1). Postings and posted labor hours show no significant change in column (2).

The second level is a panel of establishment-by-occupation units by week, with unit fixed effects in place of prefecture fixed effects and the same prefecture-level exposure, controls, and clustering. We keep units with at least four posting weeks before the shock and follow them to the end of the sample, coding weeks without postings as zero for the counts. The sample is defined by pre-shock information and excludes establishments that enter after the shock. The wage is observed only in weeks with at least one posting, and is in logs; the counts of postings and posted hours have many zeros (about 63 percent of the establishment-weeks have no posting), so we estimate them by Poisson pseudo-maximum likelihood in levels. For the wage, column (3) gives $-0.170$, smaller than the prefecture-level estimate but of the same sign and precisely estimated. The decline persists after controlling for establishment--occupation fixed effects. Changes in job composition within units and selection into posting weeks remain possible. For postings and hours the estimates are negative. Because the counts are often zero, we cannot take differences in logs in the Year over year column (4); the outcome for postings and hours is instead the growth rate of the count from the same week of the previous year, defined as the change divided by the average of the two weeks and set to zero when both counts are zero, which lies between $-2$ and $2$, and for the wage the difference in logs in the weeks in which the establishment posts in both years. The sample consists of units with at least four posting weeks in each of the two years before the shock, so that it is again defined by pre-shock information. The wage coefficient in column (4) is $-0.24$, of the same sign and similar size as column (3) ($-0.17$). The growth rates of postings and posted labor hours in column (4) show no significant change.

\begin{table}[!htbp]
  \begin{center}
      \caption{Estimates for Regions and Establishments}
      \label{tab:region_estab_did}
      \begin{tabular}[t]{lcccc}
\toprule
\multicolumn{1}{c}{ } & \multicolumn{2}{c}{Region} & \multicolumn{2}{c}{Establishment} \\
\cmidrule(l{3pt}r{3pt}){2-3} \cmidrule(l{3pt}r{3pt}){4-5}
\multicolumn{1}{c}{ } & \multicolumn{1}{c}{Level} & \multicolumn{1}{c}{Year over year} & \multicolumn{1}{c}{Level} & \multicolumn{1}{c}{Year over year} \\
\cmidrule(l{3pt}r{3pt}){2-2} \cmidrule(l{3pt}r{3pt}){3-3} \cmidrule(l{3pt}r{3pt}){4-4} \cmidrule(l{3pt}r{3pt}){5-5}
 & (1) & (2) & (3) & (4)\\
\midrule
\emph{Posted hourly wage (log)} &  &  &  & \\
Share $\times$ Post & $-0.289^{***}$ & $-0.326^{***}$ & $-0.170^{***}$ & $-0.237^{***}$\\
 & (0.070) & (0.106) & (0.023) & (0.054)\\
Observations & 9,308 & 9,294 & 2,909,487 & 658,248\\
\emph{Number of postings} &  &  &  & \\
Share $\times$ Post & $-1.013^{***}$ & $-0.857^{}$ & $-1.574^{**}$ & $-0.331^{}$\\
 & (0.377) & (1.019) & (0.676) & (0.229)\\
Observations & 9,308 & 9,294 & 7,831,641 & \vphantom{1} 4,104,864\\
\emph{Posted labor hours} &  &  &  & \\
Share $\times$ Post & $-1.295^{***}$ & $-1.018^{}$ & $-1.922^{***}$ & $-0.326^{}$\\
 & (0.496) & (1.249) & (0.498) & (0.229)\\
Observations & 9,308 & 9,294 & 7,831,641 & 4,104,864\\
\midrule
Region FE & Yes & -- & -- & --\\
Establishment-occupation FE & -- & -- & Yes & --\\
Controls & Yes & Yes & Yes & Yes\\
\bottomrule
\end{tabular}

  \end{center}\footnotesize
  \textit{Notes}: Columns (1) and (2) use 130 regions by week, with the region's 2024 Chinese share of overnight stays as Share; columns (3) and (4) use establishment-by-occupation units by week, with the prefecture's share, restricted to units with at least four posting weeks before the shock (column (4): in each of the two years before). Post equals one from the week of November 17, 2025. In the Level columns the wage is in logs (OLS), and postings and labor hours are estimated by Poisson pseudo-maximum likelihood. In the Year over year columns, the outcome is the log difference from the same week of the previous year (for establishments, the wage only) or, for establishment postings and labor hours, the growth rate (change divided by the average of the two weeks; zero when both counts are zero); the controls enter as changes from the previous year. Controls are as in Table~\ref{tab:main_did}. Standard errors clustered by region in columns (1) and (2) and by prefecture in columns (3) and (4) in parentheses. *, **, *** denote significance at 10, 5, 1 percent.
\end{table}

\subsection{Estimates by Model Specification}\label{app:by_spec}

Table~\ref{tab:main_did_by_spec} reports the estimates for all postings with the controls of the main specification added in steps: no controls in column (1), the shares of Taiwan, Korea, and the United States interacted with Post in column (2), the log minimum wage in column (3), and the Kinki-by-Expo indicator in column (4), which is the main specification and column (1) of Table~\ref{tab:main_did}. The wage estimate is $-0.11$ without any control, $-0.30$ with the other-origin controls alone, $-0.28$ when the minimum wage is added, and $-0.27$ with the Expo indicator. The controls for other origins more than double the estimate because the Chinese share is correlated with these shares and the stays of guests from these origins rose after the advisory in the prefectures that depended on Chinese guests, as Appendix~\ref{app:tourism} shows; the minimum wage and the Expo leave the estimate unchanged. For postings and posted labor hours the estimates are negative in every column and larger in size, from $-0.75$ and $-1.25$ without controls to $-1.18$ and $-1.76$ in the main specification. The 2025 revisions of the prefectural minimum wage took effect between October 2025 and early 2026, at different dates across prefectures, and column (5) adds dummies for the number of weeks since each prefecture's revision. The wage estimate falls to $-0.15$ and remains significant, posted labor hours remain significant, and postings show no significant change.

\begin{table}[!htbp]
  \begin{center}
      \caption{Difference-in-Differences Estimates by Specification, Prefecture Level}
      \label{tab:main_did_by_spec}
      \resizebox{\textwidth}{!}{\begin{tabular}[t]{lccccc}
\toprule
 & (1) & (2) & (3) & (4) & (5)\\
\midrule
\emph{Posted hourly wage (log)} &  &  &  &  & \\
Share $\times$ Post & $-0.114^{**}$ & $-0.298^{***}$ & $-0.280^{***}$ & $-0.272^{***}$ & $-0.152^{**}$\\
 & (0.044) & (0.075) & (0.078) & (0.076) & (0.067)\\
\emph{Number of postings (log)} &  &  &  &  & \\
Share $\times$ Post & $-0.748^{**}$ & $-1.315^{**}$ & $-1.313^{**}$ & $-1.180^{*}$ & $-0.802^{}$\\
 & (0.324) & (0.580) & (0.578) & (0.593) & (0.584)\\
\emph{Posted labor hours (log)} &  &  &  &  & \\
Share $\times$ Post & $-1.253^{***}$ & $-1.822^{***}$ & $-1.822^{***}$ & $-1.763^{***}$ & $-1.617^{**}$\\
 & (0.419) & (0.631) & (0.631) & (0.633) & (0.715)\\
\midrule
Prefecture FE & Yes & Yes & Yes & Yes & Yes\\
Week FE & Yes & Yes & Yes & Yes & Yes\\
Top 3 countries' shares $\times$ Post & No & Yes & Yes & Yes & Yes\\
Log minimum wage & No & No & Yes & Yes & Yes\\
Kinki $\times$ Expo period & No & No & No & Yes & Yes\\
Minimum-wage revision-week dummies & No & No & No & No & Yes\\
Observations & 3,384 & 3,384 & 3,384 & 3,384 & 3,384\\
\bottomrule
\end{tabular}
}
  \end{center}\footnotesize
  \textit{Notes}: Each coefficient is from a separate regression of the log of the outcome on Share $\times$ Post, where Share is the prefecture's 2024 Chinese share of overnight stays and Post equals one from the week of November 17, 2025. The unit is the prefecture-week (47 prefectures), and the week of November 10, 2025 is dropped. The wage is the labor-hours-weighted mean posted hourly wage. Top 3 countries are South Korea, Taiwan, and the United States; the control is each country's 2024 share of overnight stays times Post. Expo period is April 14 to October 6, 2025, and Kinki is the seven prefectures of the Kinki region. Column (4) is the main specification, column (1) of Table~\ref{tab:main_did}. Column (5) adds dummies for the number of weeks since the first revision of the prefecture's minimum wage after June 2025, binned at $-9$ and $13$ weeks. Standard errors clustered by prefecture in parentheses. *, **, *** denote significance at 10, 5, 1 percent.
\end{table}

\begin{figure}[!ht]
  \begin{center}
  \includegraphics[width = 0.8\textwidth]{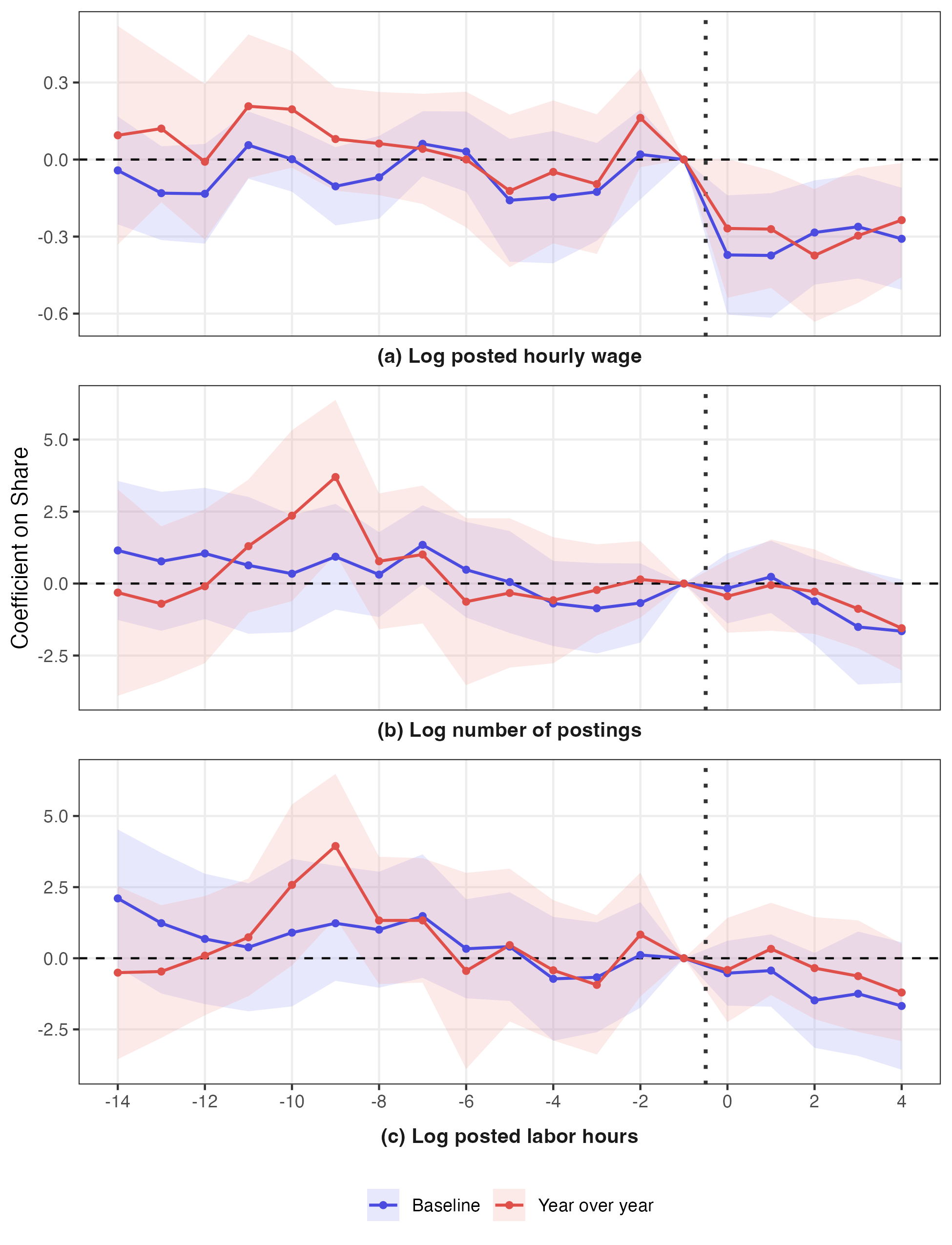}
  \caption{Event Study: Baseline and Year over Year}
  \label{fg:event_study_yoy}
  \end{center}
  \footnotesize
  \textit{Notes}: Panels show the log posted hourly wage, postings, and posted labor hours. Baseline is the specification of Figure~\ref{fg:event_study}. In Year over year, the outcome is the log difference from the same week of the previous year and the controls enter as changes from the previous year, as in column (2) of Table~\ref{tab:main_did}. The shaded area is the 90 percent confidence interval, and the block immediately before the shock is the omitted category. Standard errors are clustered by prefecture.
\end{figure}

Figure~\ref{fg:event_study_yoy} shows the event study of Section~\ref{sec:es} for the baseline specification and for the year-over-year specification of column (2) of Table~\ref{tab:main_did}. In the year-over-year specification, the outcome is the change in the log of the outcome from the same week of the previous year, and the controls enter as changes from the previous year, so that seasonal patterns and fixed differences across prefectures are removed. For the wage, the two specifications give close coefficients in every block. For postings and posted labor hours, the baseline coefficients decline toward the advisory, whereas the year-over-year coefficients show no persistent decline before it and, except for postings in the last block, no significant change after it.

\subsection{Functional Form of the Dose Response}\label{app:functional}

The exposure measure is continuous, and the three prefectures with the highest shares account for much of its range. We therefore examine whether the continuous estimate is driven by them or depends on a linear form, with two checks: excluding the prefectures with the highest shares and bins of the share.

The first check is whether the estimates are driven by the prefectures with the highest shares. Table~\ref{tab:leaveout_did} repeats the main specification without Osaka, Kyoto, and Tokyo, one at a time and all together. The wage coefficient stays close to the baseline of $-0.27$ when any one of the three prefectures is excluded, between $-0.26$ and $-0.33$, and it is $-0.34$ when all three are excluded, significant at the 1 percent level. The point estimate does not shrink when these prefectures are excluded; its standard error rises because the share then varies less. The wage result is therefore not driven by the three prefectures with the highest shares. The estimates for postings and posted labor hours are less stable: when Osaka is excluded, or when all three prefectures are excluded, posted labor hours are significant only at the 10 percent level and postings are not significant, although the point estimates are close to the baseline.

\begin{table}[!htbp]
  \begin{center}
      \caption{Estimates Excluding Prefectures with the Highest Chinese Share}
      \label{tab:leaveout_did}
      \resizebox{\textwidth}{!}{\begin{tabular}[t]{lccccc}
\toprule
 & (1) & (2) & (3) & (4) & (5)\\
\midrule
\emph{Posted hourly wage (log)} &  &  &  &  & \\
Share $\times$ Post & $-0.272^{***}$ & $-0.331^{***}$ & $-0.264^{***}$ & $-0.275^{***}$ & $-0.342^{***}$\\
 & (0.076) & (0.122) & (0.077) & (0.079) & (0.124)\\
\emph{Number of postings (log)} &  &  &  &  & \\
Share $\times$ Post & $-1.180^{*}$ & $-1.055^{}$ & $-1.138^{*}$ & $-1.125^{*}$ & $-1.099^{}$\\
 & (0.593) & (0.919) & (0.592) & (0.600) & (0.951)\\
\emph{Posted labor hours (log)} &  &  &  &  & \\
Share $\times$ Post & $-1.763^{***}$ & $-1.773^{*}$ & $-1.721^{***}$ & $-1.696^{**}$ & $-1.946^{*}$\\
 & (0.633) & (1.048) & (0.635) & (0.644) & (1.052)\\
\midrule
Excluded prefectures & None & Osaka & Kyoto & Tokyo & Osaka, Kyoto, Tokyo\\
Prefectures & 47 & 46 & 46 & 46 & 44\\
Observations & 3,384 & 3,312 & 3,312 & 3,312 & 3,168\\
\bottomrule
\end{tabular}
}
  \end{center}\footnotesize
  \textit{Notes}: Each coefficient is from a separate regression of the log of the outcome on Share $\times$ Post with the main specification (column (1) of Table~\ref{tab:main_did}); column (1) uses all 47 prefectures, and columns (2) to (5) drop the prefectures with the highest Chinese share of overnight stays. Standard errors clustered by prefecture in parentheses. *, **, *** denote significance at 10, 5, 1 percent.
\end{table}

The second check uses dummies for bins of the share, interacted with Post, to show whether the effect rises steadily with the share or levels off. Table~\ref{tab:bins_did} reports the estimates for the posted wage, postings, and posted labor hours, with bins of 1 to 2 percent, 2 to 3 percent, and 3 percent or more, relative to prefectures below 1 percent. For the wage, the decline is near zero at 1 to 2 percent, $1.1$ percent at 2 to 3 percent, and $0.8$ percent at 3 percent or more. The decline is absent at 1 to 2 percent, appears at 2 to 3 percent, and is no larger above 3 percent. The continuous estimate therefore overstates the decline in the prefectures with the highest shares: the linear fit implies $1.35$ percent for the group with 3 percent or more, against $0.8$ percent estimated.
\begin{table}[!htbp]
  \begin{center}
      \caption{Estimates by Bins of the Chinese Share of Overnight Stays}
      \label{tab:bins_did}
      \begin{tabular}[t]{lccc}
\toprule
\multicolumn{1}{c}{ } & \multicolumn{1}{c}{Posted wage} & \multicolumn{1}{c}{Postings} & \multicolumn{1}{c}{Labor hours} \\
\cmidrule(l{3pt}r{3pt}){2-2} \cmidrule(l{3pt}r{3pt}){3-3} \cmidrule(l{3pt}r{3pt}){4-4}
 & (1) & (2) & (3)\\
\midrule
Share 1--2\% $\times$ Post & $0.000^{}$ & $-0.029^{}$ & $-0.084^{***}$\\
 & (0.003) & (0.027) & (0.030)\\
Share 2--3\% $\times$ Post & $-0.011^{***}$ & $-0.057^{**}$ & $-0.081^{**}$\\
 & (0.002) & (0.024) & (0.033)\\
Share $\geq$ 3\% $\times$ Post & $-0.008^{***}$ & $-0.024^{}$ & $-0.061^{*}$\\
 & (0.003) & (0.027) & (0.032)\\
\midrule
Prefectures with share $<$ 1\% (reference) & 26 & 26 & 26\\
Prefectures with share 1--2\% & 6 & 6 & 6\\
Prefectures with share 2--3\% & 7 & 7 & 7\\
Prefectures with share $\geq$ 3\% & 8 & 8 & 8\\
Observations & 3,384 & 3,384 & 3,384\\
Controls & Yes & Yes & Yes\\
\bottomrule
\end{tabular}

  \end{center}\footnotesize
  \textit{Notes}: Each column is a separate regression of the log of the outcome (column (1): posted hourly wage; column (2): postings; column (3): posted labor hours) on dummies for the prefecture's 2024 Chinese share of overnight stays interacted with Post. Fixed effects and controls are as in Table~\ref{tab:main_did}. The reference group is prefectures with a share below 1 percent; the bins include the lower bound and exclude the upper bound. Standard errors clustered by prefecture in parentheses. *, **, *** denote significance at 10, 5, 1 percent.
\end{table}

\subsection{Differences across Shift Start Times}\label{app:band_diff}

The estimates by band in Table~\ref{tab:main_did} compare each band across prefectures, so a shock common to all bands of a prefecture, such as the revision of its minimum wage, enters each of them. Table~\ref{tab:band_diff_did} instead compares the bands within a prefecture and week. We stack the four band panels and regress the log of the band's posted wage on Share $\times$ Post $\times$ 1[early morning or daytime], with fixed effects for prefecture by week, prefecture by band, and band by week. The prefecture-by-week fixed effects absorb every shock common to the bands of a prefecture in a week, including the minimum-wage revisions, and the coefficient measures how much more the wage of early morning and daytime shifts falls than that of evening and night shifts. The controls of the main specification vary by prefecture and week and are absorbed by the prefecture-by-week fixed effects. Column (1) weights the four bands equally, and the difference is $-0.08$ and is not significant. The bands, however, differ greatly in volume: the night band holds 3 percent of postings and the daytime band 59 percent, so without weights a thin band counts as much as a thick one. Column (2) therefore weights each band by its posted labor hours, as the wage of column (1) of Table~\ref{tab:main_did} is, so that the bands are combined as in the prefecture's wage. In this specification, the wage of early morning and daytime shifts falls by 0.06 percent more per percentage point of the Chinese share than that of evening and night shifts, a difference significant at the 5 percent level.

\begin{table}[!htbp]
  \begin{center}
      \caption{Differences across Shift Start Times within Prefecture and Week}
      \label{tab:band_diff_did}
      \begin{tabular}[t]{lcc}
\toprule
  & (1) & (2)\\
\midrule
Share $\times$ Post $\times$ Early morning or daytime & $-0.080^{}$ & $-0.061^{**}$\\
 & (0.064) & (0.029)\\
\midrule
Prefecture $\times$ week FE & Yes & Yes\\
Prefecture $\times$ band FE & Yes & Yes\\
Band $\times$ week FE & Yes & Yes\\
Weighted by posted labor hours & No & Yes\\
Observations & 13,516 & 13,516\\
\bottomrule
\end{tabular}

  \end{center}\footnotesize
  \textit{Notes}: The unit is a prefecture-by-band-by-week cell, with the four shift start-time bands of Section~\ref{sec:strategy}. Each column regresses the log of the band's labor-hours-weighted mean posted hourly wage on Share $\times$ Post $\times$ 1[early morning or daytime], where Share is the prefecture's 2024 Chinese share of overnight stays and Post equals one from the week of November 17, 2025, with fixed effects for prefecture by week, prefecture by band, and band by week. Column (1) is unweighted, and column (2) weights each cell by its posted labor hours. Standard errors clustered by prefecture in parentheses. *, **, *** denote significance at 10, 5, 1 percent.
\end{table}

\subsection[Occupation Heterogeneity]{Occupation Heterogeneity}\label{app:industry}

\begin{figure}[!ht]
  \begin{center}
  \includegraphics[width = 0.95\textwidth]{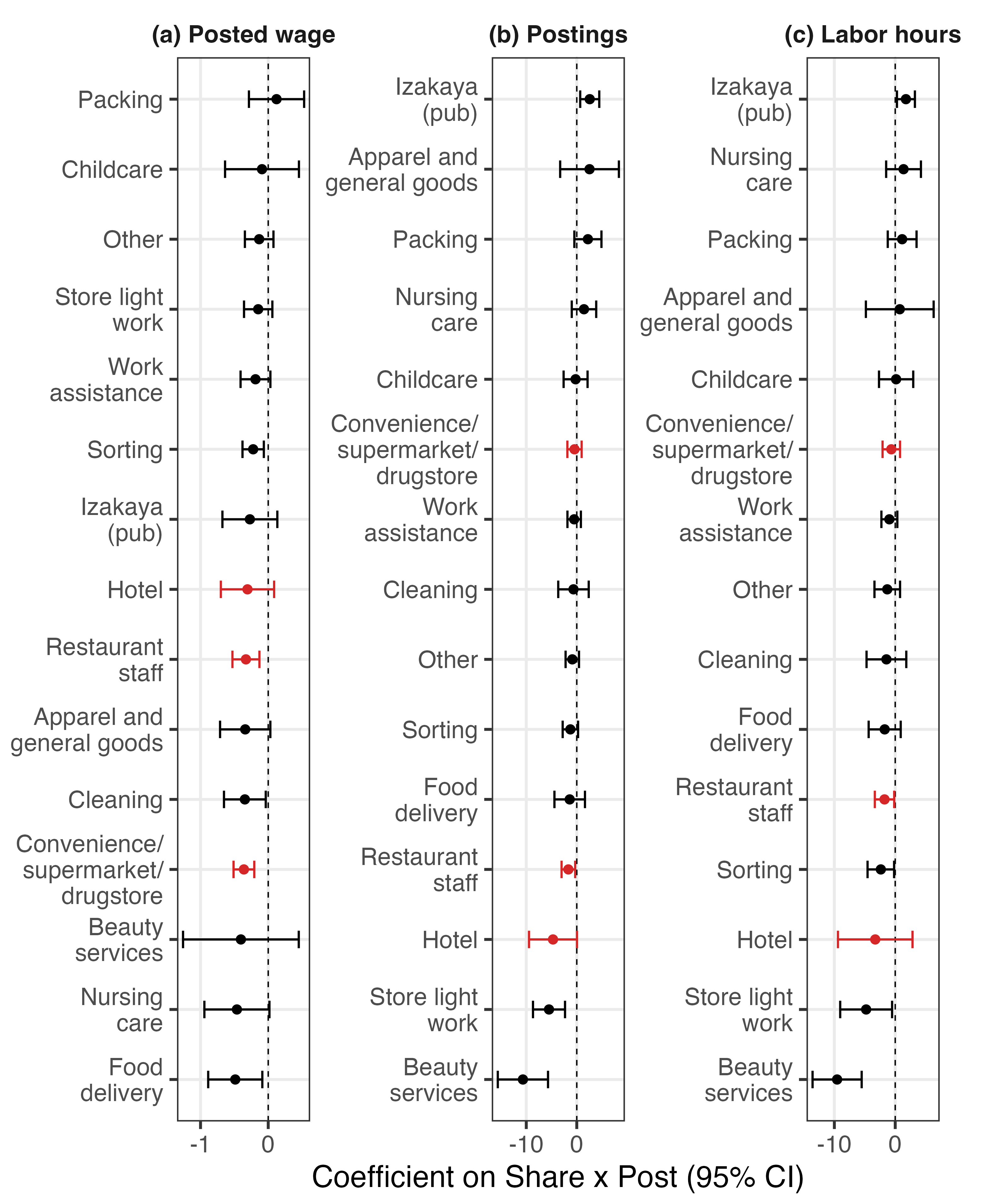}
  \caption{Response by Occupation}
  \label{fg:industry}
  \end{center}
  \footnotesize
  \textit{Notes}: Coefficient on Post $\times$ Chinese share from the main specification, estimated separately for each of the 15 occupations with the most postings: the log posted hourly wage (Panel (a)), and postings and posted labor hours by Poisson pseudo-maximum likelihood (Panels (b) and (c)). Bars are 95 percent confidence intervals clustered by prefecture. Occupations are ordered by the coefficient; hotels, restaurant staff, and convenience stores, supermarkets, and drugstores, the occupations that serve visitors most directly, are in red.
\end{figure}

The following analysis uses the platform's occupation categories. Figure~\ref{fg:industry} estimates the main specification separately for the 15 occupations with the most postings. Posted wages fall in 14 of the 15, and the decline is individually significant at the 5 percent level in five of them, including restaurant staff, convenience stores and supermarkets, and sorting work. The estimates for postings and hours differ in sign across occupations, and the largest declines are in beauty services, store light work, and hotels.

\subsection{Labor Supply Responses}\label{app:worker}

The posting data show lower wages in more exposed prefectures. This subsection uses the records of the workers who fill the shifts to ask whether incumbent workers in exposed prefectures work fewer hours after the advisory, and whether they work farther from home.

\paragraph{Sample.} We follow incumbent workers, those who completed at least three shifts in the baseline period from November 2024 to April 2025. We use the 257,275 incumbents with a registered residence, and exclude those whose median commute in the baseline exceeds 50 kilometers. For each week from May 2025 to February 2026, we compute the hours worked per incumbent, the share of incumbents who worked, the median commuting distance, and the share of hours worked outside the prefecture of residence, each aggregated to the prefecture of residence. We also follow the 36,597 tourism-attached incumbents, those who spent at least half of their baseline hours in postings whose title, description, or notes mention visitors or tourism, such as inbound tourists, hotels, inns, hot springs, or airports, and compute their hours in such postings and in all other postings, aggregated to the worker's baseline work prefecture, defined by the municipality with the most completed shifts during the baseline period.

\begin{figure}[!ht]
  \begin{center}
  \includegraphics[width = 0.95\textwidth]{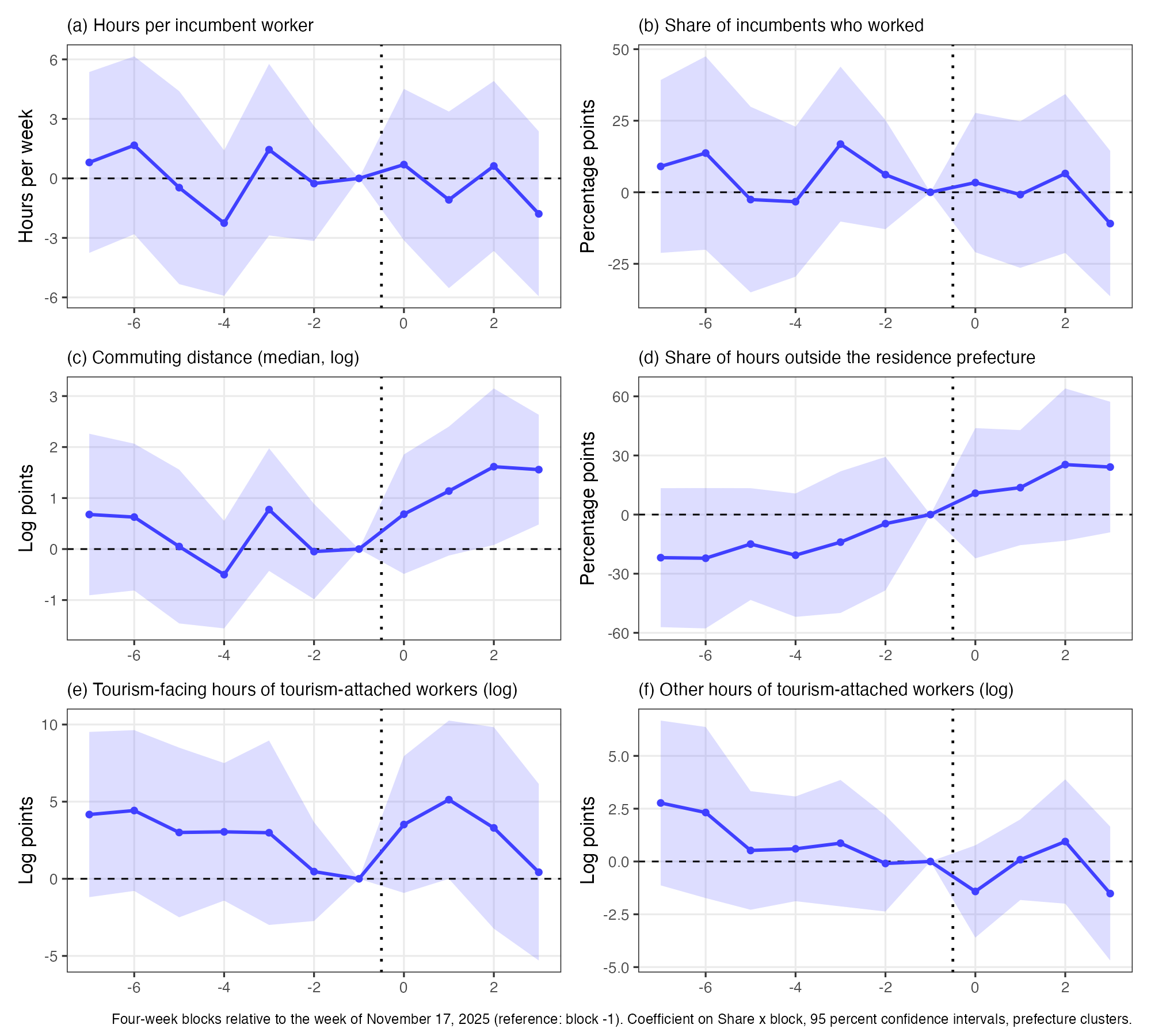}
  \caption{Event Study: Worker Mobility}
  \label{fg:worker_mobility}
  \end{center}
  \footnotesize
  \textit{Notes}: Coefficients on Share interacted with four-week blocks relative to the week of November 17, 2025, from column (2) of Table~\ref{tab:worker_mobility}. The shaded area is the 95 percent confidence interval, and the block immediately before the shock is the omitted category. Standard errors are clustered by prefecture.
\end{figure}

\paragraph{Specification.} For each outcome, we estimate the specification of equation~\eqref{eq:main} on the prefecture-by-week panel, with prefecture and week fixed effects and the controls of the main specification. Column (1) of Table~\ref{tab:worker_mobility} reports the coefficient on the Chinese share interacted with Post. Because the revisions of the prefectural minimum wage fall in the same months as the advisory and may affect where workers choose to work, column (2) adds dummies for the number of weeks since the first revision of the prefecture's minimum wage after June 2025. Figure~\ref{fg:worker_mobility} shows the event-study version of column (2) in four-week blocks. Standard errors are clustered by prefecture.

\paragraph{Results.} We find no statistically significant decline in hours worked by incumbent workers in more exposed prefectures. The hours per incumbent and the share of incumbents who worked in Panels (a) and (b) of Figure~\ref{fg:worker_mobility} stay around zero after the advisory, and the estimates in Table~\ref{tab:worker_mobility} show no significant change in either column. The confidence intervals rule out a decline in the hours per incumbent, or in the share of incumbents who worked, of more than about 1 percent of the pre-shock mean per percentage point of the Chinese share. The hours in tourism-related postings and the hours in other postings of the tourism-attached incumbents in Panels (e) and (f) show no significant change either. The tests for a pre-trend do not reject in any outcome.

Panels (c) and (d) show that the median commuting distance and the share of hours worked outside the prefecture of residence rise after the advisory in more exposed prefectures. In column (2), the median commuting distance is about 1 percent longer, and the share of hours outside the prefecture of residence about 0.3 percentage points higher, per percentage point of the Chinese share; the estimates are significant at the 5 percent level in column (2) and not significant in column (1). The share of hours outside the prefecture of residence in Panel (d) was already rising before the advisory, so the estimates point to some incumbents in exposed prefectures working farther from home but do not separate this from a trend that began earlier.

\begin{table}[!htbp]
  \begin{center}
      \caption{Worker Mobility after the Advisory}
      \label{tab:worker_mobility}
      \resizebox{\textwidth}{!}{\begin{tabular}{lccccc}
\toprule
Outcome & (1) & (2) & Pre-shock mean & Pre-trend $p$ & Observations\\
\midrule
Hours per incumbent worker & $-0.255^{}$ & $-0.378^{}$ & 3.19 & 0.39 & 1,927\\
 & (1.547) & (1.511) &  &  & \\
Share of incumbents who worked & $-3.456^{}$ & $-4.976^{}$ & 30.10 & 0.29 & 1,927\\
 & (8.646) & (8.952) &  &  & \\
Commuting distance (median, log) & $0.550^{}$ & $0.989^{**}$ & 8.61 & 0.11 & 1,927\\
 & (0.408) & (0.439) &  &  & \\
Share of hours outside the residence prefecture & $17.424^{}$ & $29.211^{**}$ & 11.85 & 0.89 & 1,927\\
 & (12.346) & (13.900) &  &  & \\
Tourism-facing hours of tourism-attached workers (log) & $0.118^{}$ & $1.309^{}$ & 1,088 & 0.58 & 1,927\\
 & (1.782) & (1.742) &  &  & \\
Other hours of tourism-attached workers (log) & $-1.164^{}$ & $-1.228^{}$ & 786 & 0.64 & 1,927\\
 & (1.058) & (1.132) &  &  & \\
\midrule
Prefecture and week fixed effects & Yes & Yes &  &  & \\
Taiwan, Korea, and USA shares $\times$ Post, log minimum wage, Kinki $\times$ Expo & Yes & Yes &  &  & \\
Minimum-wage revision relative-week dummies & No & Yes &  &  & \\
\bottomrule
\end{tabular}
}
  \end{center}\footnotesize
  \textit{Notes}: Each coefficient is from a separate regression of the outcome on Share $\times$ Post, where Share is the prefecture's 2024 Chinese share of overnight stays, on the prefecture-by-week panel from May 2025 to February 2026 (1,927 observations). The first four outcomes are for incumbent workers with a registered residence, by prefecture of residence; the last two are for tourism-attached incumbents, by baseline work prefecture, as defined above. Column (1) includes prefecture and week fixed effects and the controls of the main specification; column (2) adds dummies for the number of weeks since the first revision of the prefecture's minimum wage after June 2025. Coefficients for outcomes in logs are percent changes per percentage point of the Chinese share; the others are per unit, that is, per 100 percentage points, of the Chinese share. Pre-shock mean is the mean of the outcome before the advisory. Pre-trend $p$ is the $p$-value of a test of a linear pre-shock trend in the Chinese share. Standard errors clustered by prefecture in parentheses. *, **, *** denote significance at 10, 5, 1 percent.
\end{table}

\subsection{Placebo Tests}\label{app:placebo}

\begin{figure}[!ht]
  \begin{center}
  \includegraphics[width = 0.95\textwidth]{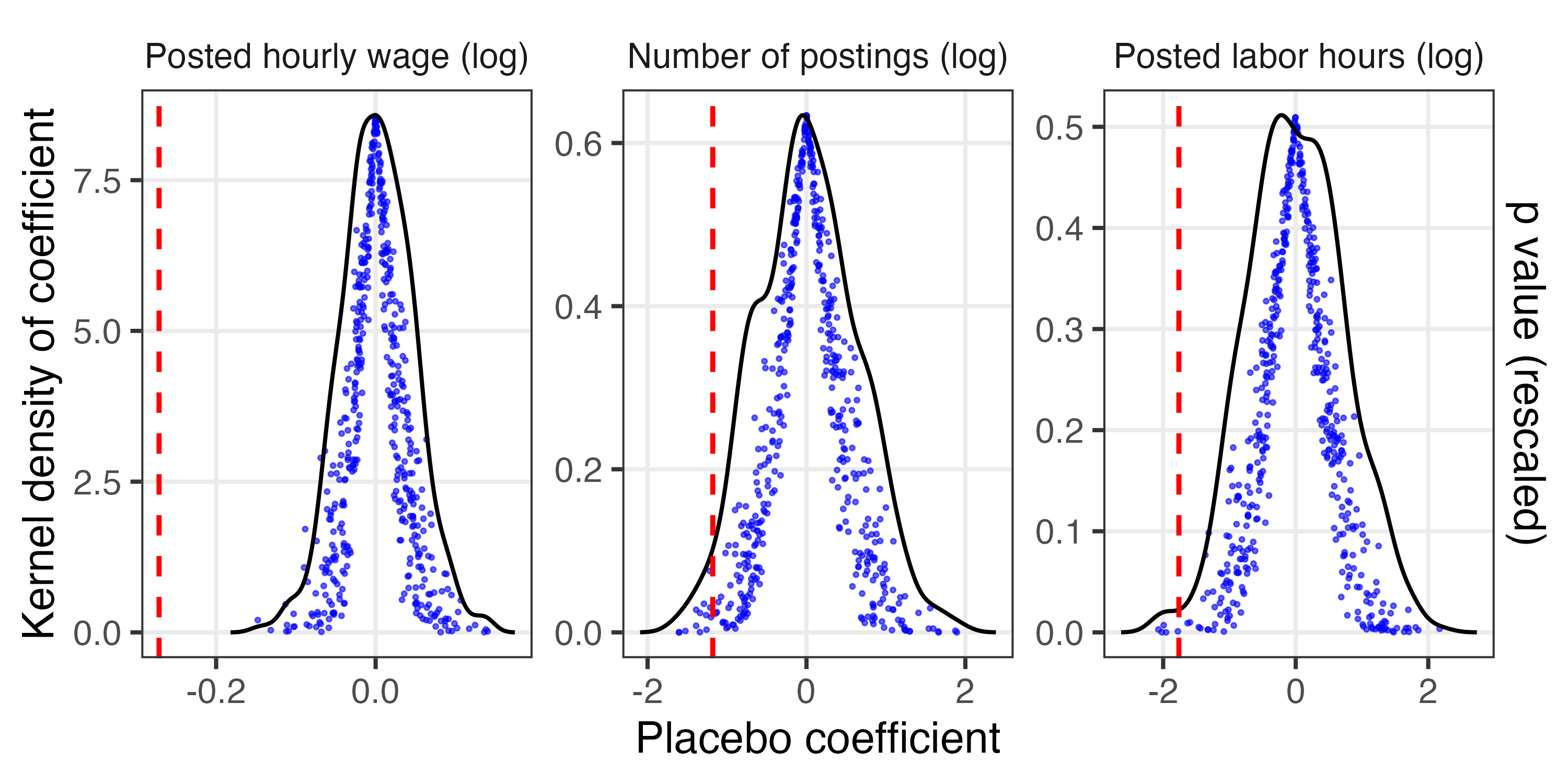}
  \caption{Mixed Placebo: Fake Shock Dates and Permuted Chinese Share}
  \label{fg:placebo_mixed}
  \end{center}
  \footnotesize
  \textit{Notes}: Kernel density of the coefficient on fake Post $\times$ Chinese share from 500 placebo regressions on the pre-shock weeks, each with a randomly drawn fake shock date and the Chinese share permuted across the 47 prefectures. Blue dots are the placebo $p$-values rescaled to the density axis, and the red dashed line is the actual coefficient from the main specification. The outcomes are the logs of the posted hourly wage (left), postings (middle), and posted labor hours (right).
\end{figure}

With 47 prefectures as clusters, we complement the clustered standard errors with the mixed placebo of \citet{chen2023mixed}. The placebo uses only the pre-shock weeks, draws a fake shock date at random, and permutes the Chinese share across the 47 prefectures, with the specification otherwise the same as the main estimate; we repeat it 500 times. Figure~\ref{fg:placebo_mixed} plots the distribution of the placebo coefficients and marks the actual coefficient, and we report the fraction of placebo coefficients that are at least as large in absolute value as the actual one. None of the 500 placebo wage coefficients is as large in absolute value as the actual estimate ($-0.27$). For postings and posted labor hours, the empirical tail frequencies are 0.062 and 0.016, so the evidence against a zero effect is weaker than for the wage. These placebos are supplementary diagnostics: interpreting their tail frequencies as valid $p$-values requires exchangeability of exposure conditional on the controls.

\subsection{Realized Decline in Chinese Stays}\label{app:iv}

The main specification uses the Chinese share in 2024 and does not use how much Chinese stays actually fell in each prefecture. We therefore measure the realized loss, for each prefecture, as the decline in Chinese overnight stays from November 2025 to March 2026 relative to the same months a year earlier, as a percentage of the prefecture's total overnight stays in those earlier months. The loss rises with the share: in the first stage, a prefecture with a one percentage point higher share loses $0.77$ percentage points more of its total stays. The share is therefore a strong predictor of the realized loss, as the first-stage statistic in Table~\ref{tab:iv_realized} shows.

We use the share interacted with Post as an instrument for the realized loss interacted with Post, with the same fixed effects, controls, and clustering as the main specification. Table~\ref{tab:iv_realized} reports the estimates. A loss of Chinese stays equal to one percent of a prefecture's total stays lowers the posted hourly wage by about $0.35$ percent, the number of postings by about $1.5$ percent, and posted labor hours by about $2.3$ percent. The ordinary least squares estimates, which use the realized loss directly, are close, at $-0.3$, $-1.9$, and $-2.6$. The two-stage estimate is the reduced-form estimate of the main specification divided by the first-stage coefficient, so it uses the same variation. Its causal interpretation additionally requires the exclusion restriction that the share affects the labor market only through the measured loss of Chinese stays. The two-stage estimate expresses the effect per unit of the realized loss rather than per unit of the share.

\begin{table}[!htbp]
  \begin{center}
      \caption{Effects per Unit of the Realized Decline in Chinese Stays}
      \label{tab:iv_realized}
      \begin{tabular}[t]{lccc}
\toprule
\multicolumn{1}{c}{ } & \multicolumn{1}{c}{Log wage} & \multicolumn{1}{c}{Log postings} & \multicolumn{1}{c}{Log labor hours} \\
\cmidrule(l{3pt}r{3pt}){2-2} \cmidrule(l{3pt}r{3pt}){3-3} \cmidrule(l{3pt}r{3pt}){4-4}
 & (1) & (2) & (3)\\
\midrule
2SLS: Loss $\times$ Post & $-0.353^{***}$ & $-1.533^{*}$ & $-2.290^{***}$\\
 & (0.108) & (0.778) & (0.827)\\
OLS: Loss $\times$ Post & $-0.255^{***}$ & $-1.861^{***}$ & $-2.626^{***}$\\
 & (0.066) & (0.645) & (0.784)\\
Reduced form: Share $\times$ Post & $-0.272^{***}$ & $-1.180^{*}$ & $-1.763^{***}$\\
 & (0.076) & (0.593) & (0.633)\\
\midrule
First stage: Share $\times$ Post on Loss & 77.0 (5.2) & 77.0 (5.2) & 77.0 (5.2)\\
First-stage $F$ statistic & 216 & 216 & 216\\
Controls & Yes & Yes & Yes\\
Observations & 3,384 & 3,384 & 3,384\\
\bottomrule
\end{tabular}

  \end{center}\footnotesize
  \textit{Notes}: Loss is the decline in Chinese overnight stays from November 2025 to March 2026 relative to the same months of the previous year, as a percentage of the prefecture's total overnight stays in those months (percentage points). The 2SLS uses Share $\times$ Post as the instrument for Loss $\times$ Post. The 2SLS and OLS coefficients are in percent per percentage point of Loss and the reduced-form coefficients in percent per percentage point of Share; the first-stage coefficient is in percentage points of Loss per unit of Share, that is, per 100 percentage points, with its standard error in parentheses. Fixed effects and controls are as in column (1) of Table~\ref{tab:main_did}. Standard errors clustered by prefecture in parentheses. *, **, *** denote significance at 10, 5, 1 percent. The 2026 values by origin are approximations based on facilities with 20 or more rooms. The scaling assumption and survey redesign are described in Appendix~\ref{app:tourism}.
\end{table}

\end{document}